# Carbon star formation as seen through the non-monotonic initial-final mass relation

Paola Marigo[1,*], Jeffrey D. Cummings[2], Jason Lee Curtis[3,4], Jason Kalirai[5,6], Yang Chen[1], Pier-Emmanuel Tremblay[7], Enrico Ramirez-Ruiz[8], Pierre Bergeron[9], Sara Bladh[1,10], Alessandro Bressan[11], Léo Girardi[12], Giada Pastorelli[1,6], Michele Trabucchi[1,13], Sihao Cheng[2], Bernhard Aringer[1], Piero Dal Tio[1,12]

*Corresponding author

[1]Department of Physics and Astronomy G. Galilei, University of Padova, Vicolo dell'Osservatorio 3, I-35122, Padova, Italy

[2]Center for Astrophysical Sciences, Johns Hopkins University, 3400 N. Charles Street, Baltimore, MD 21218, USA

[3]Department of Astrophysics, American Museum of Natural History, Central Park West, New York, NY, USA

[4]Department of Astronomy, Columbia University, 550 West 120th Street, New York, NY 10027, USA

[5]Johns Hopkins University Applied Physics Laboratory, 11101 Johns Hopkins Road, Laurel, MD 20723, USA

[6]Space Telescope Science Institute, 3700 San Martin Drive, Baltimore, MD 21218, USA

[7]Department of Physics, University of Warwick, Coventry CV4 7AL, UK

[8]Department of Astronomy and Astrophysics, University of California, Santa Cruz, CA 95064, USA

[9]*Département de Physique, Université de Montréal, C.P. 6128, Succ. Centre-Ville, Montréal, QC H3C 3J7, Canada*

[10]*Theoretical Astrophysics, Department of Physics and Astronomy, Uppsala University, Box 516, SE-751 20 Uppsala, Sweden*

[11]*International School for Advanced Studies, via Bonomea 365, I-34136 Trieste, Italy*

[12]*Astronomical Observatory of Padova – INAF, Vicolo dell'Osservatorio 5, I-35122 Padova, Italy*

[13]*Department of Astronomy, University of Geneva, Ch. des Maillettes 51, 1290 Versoix, Switzerland*

**The initial-final mass relation (IFMR) links the birth mass of a star to the mass of the compact remnant left at its death. While the relevance of the IFMR across astrophysics is universally acknowledged, not all of its fine details have yet been resolved. A new analysis of a few carbon-oxygen white dwarfs in old open clusters of the Milky Way led us to identify a kink in the IFMR, located over a range of initial masses, $1.65 \lesssim M_i/M_\odot \lesssim 2.10$. The kink's peak in WD mass of ≈ 0.70 – 0.75 $M_\odot$ is produced by stars with $M_i$ ≈ 1.8 – 1.9 $M_\odot$, corresponding to ages of about 1.8 – 1.7 Gyr. Interestingly, this peak coincides with the initial mass limit between low-mass stars that develop a degenerate helium core after central hydrogen exhaustion, and intermediate-mass stars that avoid electron degeneracy. We interpret the IFMR kink as the signature of carbon star formation in the Milky Way. This finding is critical to constraining the evolution and chemical enrichment of low-mass stars, and their impact on the spectrophotometric properties of galaxies.**

Carbon is essential to life on Earth, but its origin in the Milky Way (MW) is still debated: some studies[1,2] place the major site of its synthesis in the winds of massive stars that eventually exploded as supernovae, others[3,4] are in favour of low-mass stars that blew off their envelopes by stellar winds and became white dwarfs (WDs). Within the latter group, the primary sources of carbon are the winds of carbon stars, characterised by a photospheric carbon-to-oxygen ratio C/O > 1, which form during the thermally-pulsing asymptotic giant branch (TP-AGB) phase as a consequence of repeated third dredge-up (3DU) episodes. So far, the range of initial masses of carbon stars and their chemical ejecta are not accurately known from theory[5] since these quantities depend on a number of complex physical processes that are difficult to model, convection and mass loss above all.

Here we show that the initial-final mass relation (IFMR) of white dwarfs may help shed light on this matter. The IFMR connects the mass of a star on the main sequence, $M_i$, with the mass, $M_f$, of the WD left at the end of its evolution. This fate[6] is common to low- and intermediate-mass stars ($0.9 \lesssim M_i/M_\odot \lesssim 6$–$7$) that, after the exhaustion of helium in the core, go through the AGB phase and produce carbon-oxygen WDs. The IFMR is also useful to investigate the quasi-massive stars ($8 \lesssim M_i/M_\odot \lesssim 10$) that, after the carbon burning phase, evolve as super-AGB stars and eventually produce oxygen–neon–magnesium WDs.

The IFMR plays a key role in several fields of modern astrophysics[7]. Once $M_i$ is known, the IFMR fixes the mass of the metal-enriched gas returned to the interstellar medium, thus putting constraints to the efficiency of stellar winds during the previous evolution. The high-mass end of the IFMR provides an empirical test to determine the maximum initial mass for stars that develop degenerate oxygen–neon–magnesium cores and proceed through the super-AGB phase without exploding as electron-capture supernovæ. The IFMR is also relevant in a wider framework as a key ingredient in chemical evolution models of galaxies; at the same time it sets a lower limit to the nuclear fuel burnt during the TP-AGB, therefore constraining the contribution of this phase to the integrated light of galaxies[8–10].

To derive the semi-empirical IFMR, singly-evolved WDs that are members of star clusters are ideally used[11–13]. Spectroscopic analysis provides their atmospheric parameters, that is, surface gravity, effective temperature, and chemical composition. Coupling this information to appropriate WD cooling models provides the WD mass, its cooling age, and additional parameters for testing cluster membership and single-star status. Finally, subtracting a WD's cooling age from its cluster's age gives the evolutionary lifetime of its progenitor, and hence its $M_i$.

Previous analyses[12,14] of old open clusters with ages $\gtrsim$ 1.5 Gyr (NGC 6121, NGC 6819, and NGC 7789) explored WDs that recently evolved from stars with $M_i$ < 2 $M_\odot$, and this showed possible signatures of non-linearity in the IFMR near $M_i \sim$ 2 $M_\odot$. Updated analyses of these 12 WDs using improved models and uniformly analysed cluster parameters[15], and the addition of 7 WDs in the old open cluster M67[16,17], further supported that the low-mass IFMR is nonlinear and potentially not even monotonically increasing.

# 1 The IFMR kink

Compared with the previous studies just mentioned, this study introduces two novel elements: (1) the discovery of seven WD members of the old open clusters NGC 752 (age ≃ 1.55 Gyr) and Ruprecht 147 (age ≃ 2.5 Gyr), and (2) the use of a new analysis technique that for the first time combines photometric and spectroscopic data to better constrain the WD parameters. The same technique is applied to both the 7 newly discovered WDs and to reanalyse 12 previously published WDs with $M_i$ < $2.1 M_\odot$ (see Table 1). Our new methodology is detailed and discussed in Methods. These data establish a low-mass IFMR kink across the interval $1.65 \lesssim M_i/M_\odot \lesssim 2.10$, as shown in Fig. 1. All clusters involved in the IFMR kink have near-solar metallicity, $-0.1 \lesssim [Fe/H] \lesssim 0.1$.

While the steep increase in the IFMR near $M_i \simeq 1.65 M_\odot$ is at present well constrained, further data are needed to better probe the decreasing IFMR for $M_i \gtrsim 2 M_\odot$. However, we underline that after this rapid rise such a temporary decrease is necessary to keep consistency with the observed field-WD mass distribution. For example, if this steep rise was instead followed by a plateau at $M_f$ ~0.7 $M_\odot$, then every progenitor with $1.9 \lesssim M_i/M_\odot \lesssim 2.8$ would create a ~ 0.7 $M_\odot$ WD, substantially overproducing field WDs at this mass compared with observations[15,18]. This aspect is thoroughly discussed in the Supplementary Section "The WD mass distribution" (see also Supplementary Figs. 1-2). The kink also illustrates a limitation of inferring the IFMR directly from such field WD mass distributions because a monotonic form must be assumed, for example, using the Gaia Data Release 2 (DR2)[19].

| Name | RA [h:min:s] | DEC [° : ' : "] | $T_{\text{eff}_S}$ [K] | $\log(g)_S$ [cm s$^{-2}$] | $M_{f_S}$ [M$_\odot$] | $M_G$ [mag] | $M_{f_P}$ [M$_\odot$] | $M_f$ [M$_\odot$] | $\tau$ [Myr] | $M_i$ [M$_\odot$] |
|---|---|---|---|---|---|---|---|---|---|---|
| **DA Members** | | | | | | | | | | |
| R147-WD01 | 19:15:33.8 | -16:52:49 | 17850±250 | 8.11±0.04 | 0.69±0.03 | 11.08±0.05 | 0.66±0.02 | 0.67±0.02 | $129^{+11}_{-9}$ | $1.66^{+0.01}_{-0.01}$ |
| R147-WD04 | 19:16:59.7 | -16:31:21 | 19550±250 | 8.07±0.04 | 0.66±0.03 | 10.93±0.05 | 0.67±0.02 | 0.67±0.02 | $92^{+10}_{-8}$ | $1.65^{+0.01}_{-0.01}$ |
| R147-WD07 | 19:16:13.7 | -16:20:13 | 15900±200 | 8.08±0.04 | 0.66±0.03 | 11.35±0.05 | 0.68±0.02 | 0.67±0.02 | $193^{+13}_{-11}$ | $1.68^{+0.01}_{-0.01}$ |
| R147-WD08 | 19:18:44.3 | -15:53:56 | 13000±200 | 8.11±0.05 | 0.67±0.03 | 11.78±0.05 | 0.70±0.03 | 0.69±0.02 | $368^{+22}_{-21}$ | $1.73^{+0.01}_{-0.01}$ |
| R147-WD10 | 19:18:01.7 | -15:49:56 | 17950±250 | 8.08±0.04 | 0.67±0.03 | 11.14±0.05 | 0.69±0.02 | 0.68±0.02 | $131^{+11}_{-9}$ | $1.66^{+0.01}_{-0.01}$ |
| N752-WD01 | 01:59:05.5 | +38:03:38 | 15500±250 | 8.12±0.04 | 0.69±0.03 | 11.26±0.05 | 0.62±0.02 | 0.65±0.02 | $191^{+15}_{-14}$ | $2.10^{+0.02}_{-0.02}$ |
| **DB Member** | | | | | | | | | | |
| R147-WD02 | 19:12:52.5 | -16:14:35 | 16300±200 | 8.12±0.05 | 0.67±0.03 | 11.23±0.05 | 0.66±0.02 | 0.66±0.02 | $188^{+13}_{-10}$ | $1.68^{+0.01}_{-0.01}$ |
| **Non-members** | | | | | | | | | | |
| R147-WD03 | 19:18:58.2 | -16:48:22 | 15300±250 | 8.00±0.04 | 0.61±0.03 | – | – | – | $185^{+16}_{-15}$ | – |
| R147-WD09 | 19:15:03.7 | -16:03:41 | 15800±200 | 7.96±0.04 | 0.59±0.02 | – | – | – | $155^{+14}_{-13}$ | – |
| R147-WD11 | 19:14:38.1 | -15:58:58 | 14900±200 | 8.00±0.04 | 0.61±0.03 | – | – | – | $202^{+17}_{-16}$ | – |
| R147-WD15 | 19:15:18.2 | -17:37:10 | – | – | – | – | – | – | – | – |

| Name | RA [h:min:s] | DEC [° : ' : "] | $T_{\text{eff}_S}$ [K] | $\log(g)_S$ [cm s$^{-2}$] | $M_{f_S}$ [M$_\odot$] | $M_V$ [mag] | $M_{f_P}$ [M$_\odot$] | $M_f$ [M$_\odot$] | $\tau$ [Myr] | $M_i$ [M$_\odot$] |
|---|---|---|---|---|---|---|---|---|---|---|
| **Previous Low-Mass WD Parameters Updated with Photometric Analysis** | | | | | | | | | | |
| M4-WD00 | 16:23:49.9 | -26:33:32 | 20900±500 | 7.77±0.08 | 0.51±0.04 | 10.57±0.15 | 0.59±0.06 | 0.53±0.03 | $37^{+6}_{-4}$ | $0.87^{+0.01}_{-0.01}$ |
| M4-WD04 | 16:23:51.3 | -26:33:04 | 25450±550 | 7.78±0.07 | 0.52±0.03 | 9.94±0.15 | 0.51±0.05 | 0.52±0.03 | $16^{+1}_{-1}$ | $0.87^{+0.01}_{-0.01}$ |
| M4-WD05 | 16:23:41.4 | -26:32:53 | 28850±500 | 7.77±0.07 | 0.53±0.03 | 9.96±0.15 | 0.61±0.06 | 0.55±0.03 | $10^{+1}_{-1}$ | $0.87^{+0.01}_{-0.01}$ |
| M4-WD06 | 16:23:42.3 | -26:32:39 | 26350±500 | 7.90±0.07 | 0.59±0.04 | 9.90±0.15 | 0.52±0.05 | 0.56±0.03 | $14^{+2}_{-1}$ | $0.87^{+0.01}_{-0.01}$ |
| M4-WD15 | 16:23:51.0 | -26:31:08 | 24600±600 | 7.89±0.08 | 0.57±0.04 | 9.98±0.15 | 0.50±0.05 | 0.54±0.03 | $18^{+3}_{-2}$ | $0.87^{+0.01}_{-0.01}$ |
| M4-WD20 | 16:23:46.5 | -26:30:32 | 21050±550 | 7.79±0.08 | 0.52±0.04 | 10.26±0.15 | 0.49±0.05 | 0.50±0.03 | $33^{+5}_{-3}$ | $0.87^{+0.01}_{-0.01}$ |
| M4-WD24 | 16:23:41.2 | -26:29:54 | 26250±500 | 7.79±0.07 | 0.53±0.03 | 9.97±0.15 | 0.55±0.05 | 0.54±0.03 | $14^{+1}_{-1}$ | $0.87^{+0.01}_{-0.01}$ |
| N6819-6 | 19:41:20.0 | +40:02:56 | 21700±350 | 7.94±0.05 | 0.60±0.03 | 10.49±0.10 | 0.58±0.04 | 0.59±0.02 | $40^{+8}_{-5}$ | $1.61^{+0.01}_{-0.01}$ |
| N7789-5 | 23:56:49.1 | +56:40:13 | 31700±450 | 8.12±0.06 | 0.71±0.04 | 9.96±0.10 | 0.69±0.04 | 0.70±0.03 | $8^{+1}_{-1}$ | $1.90^{+0.01}_{-0.01}$ |
| N7789-8 | 23:56:57.2 | +56:40:01 | 24800±550 | 8.11±0.07 | 0.70±0.04 | 10.62±0.10 | 0.73±0.04 | 0.71±0.03 | $35^{+8}_{-5}$ | $1.91^{+0.01}_{-0.01}$ |
| N7789-11 | 23:56:30.8 | +56:37:19 | 20500±650 | 8.27±0.10 | 0.79±0.06 | 10.83±0.10 | 0.67±0.04 | 0.70±0.03 | $81^{+16}_{-15}$ | $1.94^{+0.02}_{-0.02}$ |
| N7789-14 | 23:56:37.8 | +56:39:08 | 21100±950 | 7.99±0.14 | 0.62±0.08 | 11.02±0.10 | 0.77±0.04 | 0.74±0.04 | $85^{+21}_{-18}$ | $1.94^{+0.02}_{-0.02}$ |

Table 1: Main parameters of the WDs and their stellar progenitors. The parameters for the spectroscopically observed Ruprecht 147 and NGC 752 WD candidates are organised by atmospheric composition (DA/DB) and membership, including the apparent background star R147-WD15. Updated parameters of previously published low-mass WDs are also given. Columns from left to right are WD name, right ascension, declination, spectroscopic-based effective temperature, logarithm of spectroscopic-based surface gravity, spectroscopic-based final mass, photometric-based absolute magnitude, photometric-based final mass, weighted-average final mass, weighted-average WD cooling age, progenitor's initial mass. The spectroscopic-based errors are from fitting and external errors. The photometric-based errors are from photometric and cluster-parameter errors. For stars classified as non-members of Ruprecht 147, cluster-related parameters are unknown and hence appear as empty cells in the table. The background star R147-WD15 was recognized not to be a WD and therefore no spectroscopic parameter was derived.

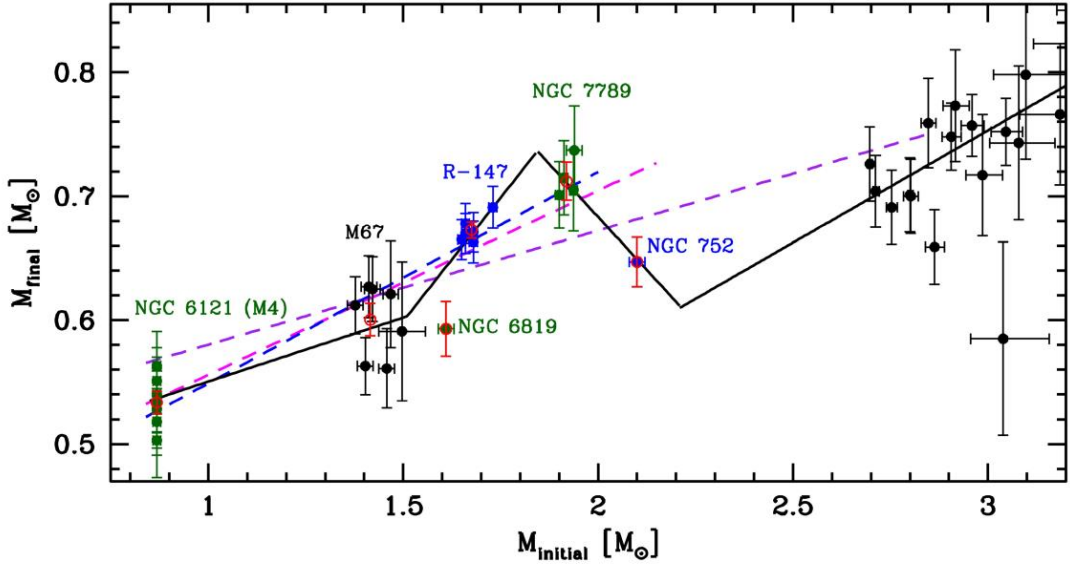

Figure 1: The semi-empirical IFMR. The data includes the 7 newly discovered WDs (blue data) and the 12 previously published WDs (green data) reanalysed with a method that couples spectroscopy, photometry, and the Gaia-based distances of the host star clusters (see Methods for more details). Additionally, previous intermediate-mass data and the M67 data are shown in black. Error bars cover a range of $\pm 1\,\sigma$ (see Table 1). Each WD group's cluster is also labelled and each of their parameter averages is overlaid in open red data points. Based on these IFMR data and their $1\sigma$ error, we illustrate three weighted linear fits, namely: from $M_i$ of 0.85 $M_\odot$ to 2.85 $M_\odot$ (purple), from $M_i$ of 0.85 $M_\odot$ to 2.15 $M_\odot$ (magenta), and from $M_i$ of 0.85 $M_\odot$ to 2.00 $M_\odot$ (blue). As discussed in Methods, these three fits provide poor representations of the observed data. Conversely, the 4-piece linear fit (black), which draws a kink in $M_f$ around $M_i \simeq 1.8\,M_\odot$, does provide a strong representation of the data and is our observational IFMR.

## 2  Physical interpretation

We interpret the kink in the IFMR as the signature of the lowest-mass stars in the MW that became carbon stars during the TP-AGB phase. The inflection point near $1.65 M_\odot$ (Fig. 1) should mark the minimum initial mass for a solar metallicity star to become a carbon star as a consequence of the 3DU episodes during thermal pulses (TP).

The proposed explanation is as follows. At solar-like metallicity low-mass carbon stars ($1.65 \lesssim M_i/M_\odot \lesssim 1.90$) attain low C/O ratios ($\lesssim 1.3$) and low values of the excess of carbon compared to oxygen, $C - O$, in the atmosphere. The quantity $C - O$ is particularly relevant as it measures the budget of free carbon, not locked in the CO molecule, available to condense into dust grains. In fact, state-of-the-art dynamical models for carbon stars[20–22] predict that a minimum carbon excess, $(C - O)_{min}$, is necessary to generate dust-driven winds, with mass-loss rates exceeding a few $10^{-7}\,M_\odot\,yr^{-1}$. We recall that according to a standard notation, $C - O = \log(n_C - n_O) - \log(n_H) + 12$, where $n_C$, $n_O$, and $n_H$ denote the number densities of carbon, oxygen, and hydrogen, respectively. More details about the wind models for carbon stars, $(C - O)_{min}$, and our CDYN prescription for mass loss are provided in Methods.

The existence of a threshold in carbon excess impacts on the TP-AGB evolution and hence on the IFMR. In TP-AGB stars the surface enrichment of carbon is controlled by the 3DU, a series of mixing episodes that happen each time the base of the convective envelope is able to penetrate into the inter-shell region left at the quenching of a thermal pulse[6]. The efficiency of a 3DU event is commonly described by the dimensionless parameter $\lambda = \Delta M_{3DU}/\Delta M_c$, defined as the amount of dredged-up material, $\Delta M_{3DU}$, relative to the growth of the core mass, $\Delta M_c$, during the previous inter-pulse period.

Figure 2 illustrates how the efficiency of 3DU regulates the increase of the surface C/O and hence the carbon excess; how the latter, in turn, affects the mass-loss rate, hence the lifetime of a carbon star and eventually the final mass of the WD. The models refer to a star with $M_i = 1.8 M_\odot$ and $Z = 0.014$, near the kink's peak of the semi-empirical IFMR. The two cases in Fig. 2a,b share the same set of input prescriptions, except that the 3DU is shallow in the model of Fig. 2a ($\lambda \simeq 0.17$), and much more efficient in the model of Fig. 2b ($\lambda = 0.5$ as the star becomes C-rich). In both cases, as soon as the star reaches C/O > 1, a sudden drop in the mass-loss rate is expected to occur.

This prediction deserves to be explained in detail. The transition from C/O < 1 to C/O > 1 marks a radical change both in the molecular abundance pattern of the atmosphere (shifting from O-bearing to C-bearing species)[23], and in the mineralogy of the dust that could in principle condense in the coolest layers.

When a carbon star is born, the silicate-type dust that characterises the circumstellar envelopes of M-type stars is no longer produced and the composition of the grains that may actually form suddenly changes, switching mainly to silicon carbide and amorphous carbon[24–27]. The key point is that the growth of carbonaceous dust requires suitable physical conditions[25,28] (e.g., temperature, density and chemical composition of the gas, stellar radiation field), and these may not always be fulfilled as soon as C/O $\gtrsim$ 1. In addition, carbonaceous grains are expected to drive a wind only if they form in sufficient amount[20–22], a condition expressed by the threshold $(C-O)_{\min}$.

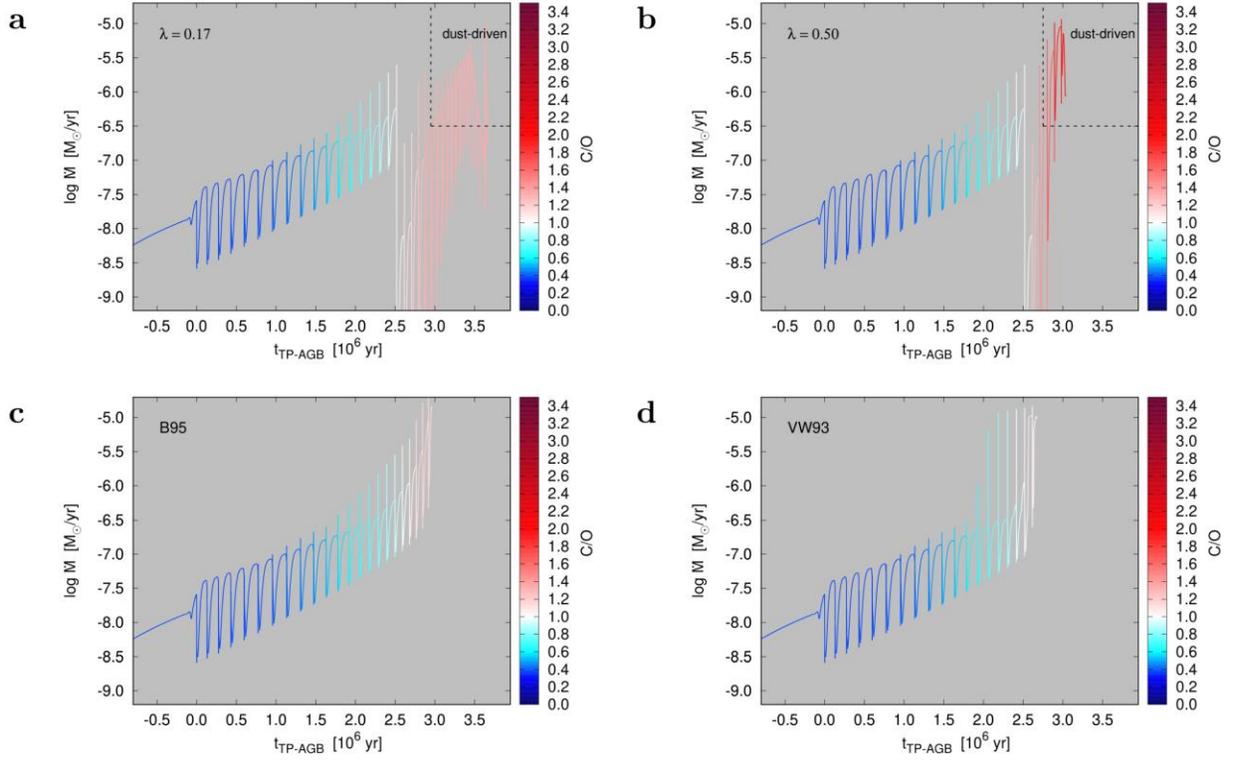

Figure 2: Evolution of the mass-loss rate during the whole TP-AGB evolution of a star with $M_i = 1.8\,\mathrm{M_\odot}$ and solar metallicity. Time is set to zero at the first TP. The tracks are colour-coded according to the current photospheric C/O ratio. Calculations differ in the treatment of mass loss and/or in the efficiency $\lambda$ of the 3DU. **a - b**, Models in which mass loss during the carbon star phase does depend on the carbon excess (CDYN prescription; see Methods). The 3DU is less efficient in model **a** compared to model **b**. Inside the region delimited by the dotted lines stellar winds are driven by carbonaceous dust grains. **c - d**, Results obtained with mass-loss formalisms that, unlike in **a** and **b**, do not contain an explicit dependence on the carbon abundance. See Methods for more details.

It follows that, as long as the atmospheric abundance of free carbon is small, typically during the early carbon-star stages, dust grains cannot be abundantly produced. This circumstance is clearly shown by computations[24] of dust-grain growth as a function of C/O and mass-loss rate.

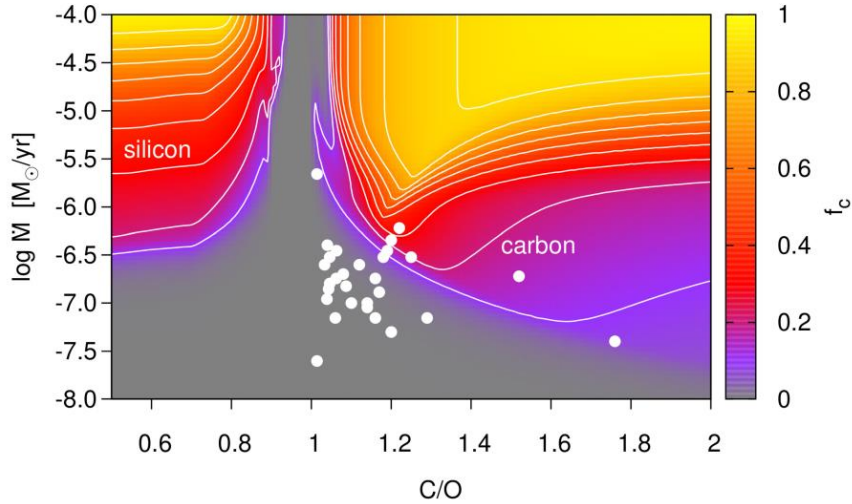

Figure 3: Map of the condensation factor, $f_c$, as a function of mass-loss rate and photospheric C/O. The assumed stellar parameters, $M = 1.5\,M_\odot$, $\log(L/L_\odot) = 3.85$, $T_{\rm eff} = 2950\,{\rm K}$, are extracted from the TP-AGB track (just before the $18^{\rm th}$ TP) of a model with $M_{\rm i} = 1.8\,M_\odot$ and $Z = 0.014$. The quantity $f_c$ refers to the fraction of silicon that condenses in silicates (olivine, pyroxene and quartz) for $C/O < 1$, while it corresponds to the condensed fraction of free carbon for $C/O > 1$. White dots show the location of a sample[29] of carbon-rich SRVs in the MW with measured mass-loss rates and C/O. According to the map, most of these carbon stars should be dust free, in full agreement with their observed spectra[30].

As we see in Fig. 3 the condensation factor of carbon, $f_c$, is extremely low for C/O in a narrow range just above unity, irrespective of the mass-loss rate. The inefficiency of the carbon dust condensation extends to larger C/O for low mass-loss rates ($\dot{M} \lesssim 3 \times 10^{-7}$ $M_\odot$ yr$^{-1}$). It should be kept in mind that the map of $f_c$ corresponds to a pre-built grid of ($\dot{M}$, C/O) combinations, and clearly not all of them represent realistic cases.

At the same time such a map is very useful when analysing observed combinations of ($\dot{M}$, C/O), such as those referring to a sample[29] of carbon-rich irregular and semi-regular variables (SRV). According to our interpretation these variables should represent the progenitors of the WDs that populate the IFMR kink. For an extensive discussion about this point, see the section entitled "Other supporting evidence: Galactic semi-regular variables" in the Supplementary Information (see also Supplementary Figs. 4-6).

In Fig. 3 most of the observed SRV fall in a region of low mass-loss rate and small carbon enrichment, for which we expect an inefficient dust condensation of carbon grains. The natural conclusion is that if carbonaceous dust grains are not abundant enough to drive a powerful wind, when C/O just exceeds unity, then only a modest outflow may be generated, possibly sustained by small-amplitude pulsations, like those of SRV stars[31,32]. These conditions apply to the model of Fig. 2a, which experiences a shallow 3DU. The C/O ratio grows slowly (maximum of $\simeq 1.33$), the star stays in a phase of very low mass loss until the threshold in carbon excess is slightly overcome and a moderate dust-driven wind is eventually activated, with mass-loss rates not exceeding few $10^{-6}$ $M_\odot$ yr$^{-1}$. Also the model shown in Fig. 2b enters a phase of low mass loss soon after the transition to carbon star, but then its evolution proceeds differently. As the 3DU is more efficient, C/O increases more rapidly (maximum of $\simeq 1.91$) so that the threshold in carbon excess is largely overcome, a powerful dust-driven wind is generated, and the mass-loss rate rises up to $\simeq 10^{-5}\,M_\odot\,\mathrm{yr}^{-1}$.

These model differences affect the carbon star lifetimes and, in turn, the final masses left after the TP-AGB phase. In the model of Fig. 2a the carbon star phase lasts $\simeq 1.15$ Myr and produces a WD with a final mass of $\simeq 0.732 M_\odot$. In model of Fig. 2b the duration of the carbon star phase is halved, $\simeq 0.52$ Myr, and terminates with a WD mass of $\simeq 0.635 M_\odot$.

Likewise, shorter lifetimes and lower final masses are obtained if we adopt mass-loss formulations that do not depend on the carbon abundance. In Fig. 2c,d we show two examples in which we use the B95 and VW93 relations. Both predict a systematic increase of the average mass-loss rate as the star evolves on the TP-AGB. The resulting WD masses are $0.646 M_\odot$ and $0.615 M$, respectively.

## 3  The IFMR and carbon star formation in the MW

While the above considerations, relative to a given TP-AGB model, provide the physical key to interpret the data, we aim to build up an overall picture as a function of the initial mass of the star. Therefore, we employ the large model grid introduced in Methods, which consists of TP-AGB calculations that cover a relevant region of the parameter space ($M_i$, $\lambda$), assuming solar initial metallicity.

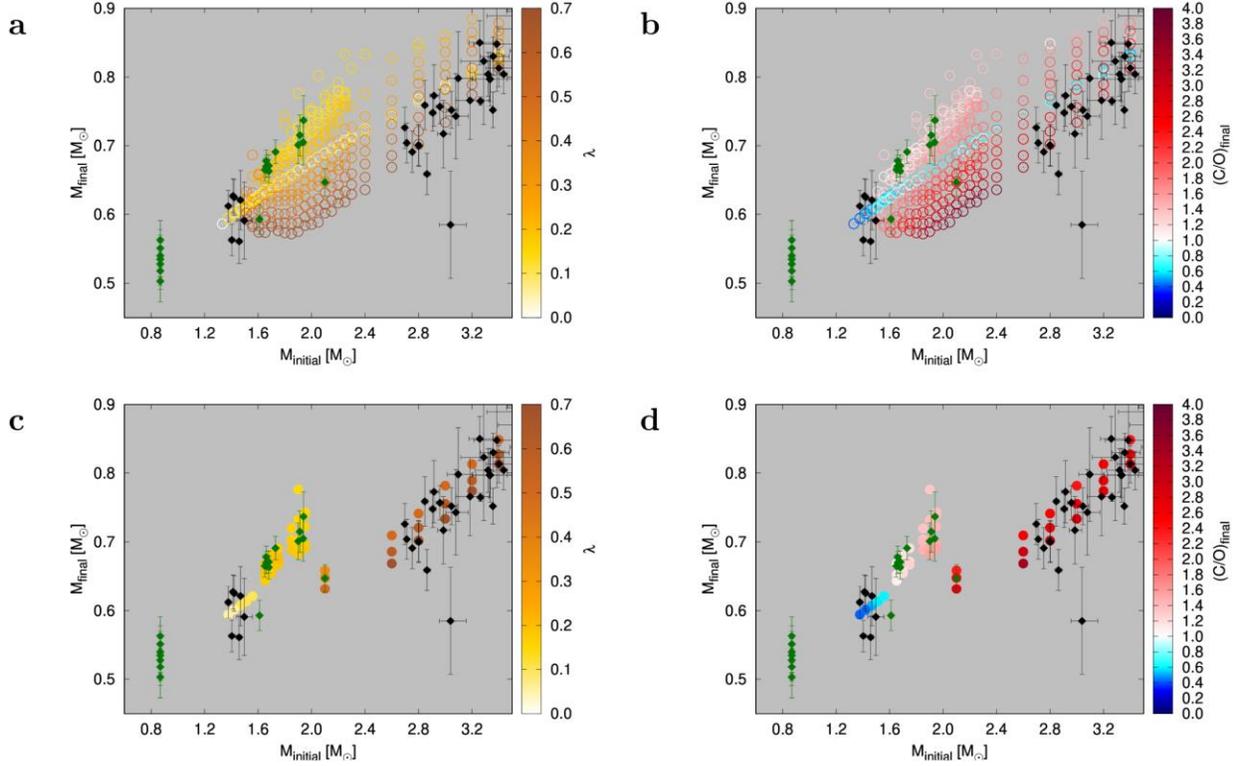

Extended Data Figure 1: Comparison between the semi-empirical IFMR and model results. The semi-empirical data are shown with diamonds and error bars covering a range of $\pm 1\,\sigma$. Newly discovered and newly analysed WD data (see Table 1) are shown in green. **a** - **b**, Predictions for the whole $(M_\mathrm{i}, \lambda)$ grid of models. **c** - **d**, Selected models that are found to match the semi-empirical IFMR. The theoretical IFMR is colour-coded according to the values of the efficiency of the 3DU (**a** - **c**) and the photospheric C/O at the end of the TP-AGB phase (**b** - **d**).

The results for the entire grid are displayed in Extended Data Fig. 1a,b. We see that varying $\lambda$ at a given $M_i$ produces a large dispersion in the final WD mass and C/O. In particular, $M_f$ of carbon stars anticorrelates with $\lambda$. At this point, the natural step is to pick up the ($M_i$, $\lambda$) combinations that best approximate the semi-empirical IFMR (Extended Data Figs. 1c,d).

The detected IFMR kink over the range $1.65 \lesssim M_i/M_\odot \lesssim 2.0$ is well recovered by assuming that these stars experience a shallow 3DU during the TP-AGB phase, typically with $0.1 \lesssim \lambda \lesssim 0.2$ (Extended Data Fig. 1c). Stars in the low-mass part ($1.65 \lesssim M_i/M_\odot < 1.8$) are those just massive enough to become carbon stars (Extended Data Fig. 1d). They are little enriched in carbon, with low final ratios ($1.1 \lesssim C/O \lesssim 1.2$) and low carbon excesses ($7.55 \lesssim C - O \lesssim 8.10$). In these models, carbonaceous dust grains are not expected to form in sufficient amount to trigger a vigorous radiation-driven wind[21,33], while pulsation is likely to strip these stars of their small envelopes. The peak at $M_i \simeq 1.8$–$1.9 M_\odot$ corresponds to stars that reach a final $C/O \approx 1.3$ when they overcome the minimum threshold, $(C - O)_{min} \simeq 8.2$, late, close to the end of their evolution, and when the core has already grown appreciably in mass ($M_f \gtrsim 0.7 M_\odot$). Beyond the peak, at larger initial masses ($1.95 \lesssim M_i/M_\odot \lesssim 2.1$), the IFMR data are matched with TP-AGB models having a moderately higher efficiency, $0.2 \lesssim \lambda \lesssim 0.4$ (Extended Data Fig. 1c). Carbon enrichment is therefore somewhat larger and this leads to an earlier activation of the dust-driven wind. As a consequence, the WD masses are predicted to decrease until the IFMR regains a positive slope, as shown by the data for $M_i \geq 2.8 M_\odot$ (Extended Data Fig. 1d). Models that reproduce this linear portion of the IFMR, with $2.6 \lesssim M_i/M_\odot \lesssim 3.5$, are characterised by a fairly higher efficiency of the 3DU, $0.5 \lesssim \lambda \lesssim 0.7$.

They reach larger values of the final carbon excess ($8.7 \lesssim C - O \lesssim 9.2$) and carbon-to-oxygen ratio ($2 \lesssim C/O \lesssim 4$) so that dust grains are expected to condense plentifully in their extended atmospheres[21,22].

Using the ranges of $\lambda$ just obtained, we may constrain the average efficiency of the 3DU as a function of $M_i$ and derive the relation shown in Fig. 4a, (orange curve). Adopting such a relation in our TP-AGB calculations, we get a theoretical IFMR that recovers quite well the semi-empirical data (Fig. 4b), in particular the peak around $\simeq 1.8 - 1.9 M_\odot$. We emphasise that the proposed calibration of the 3DU at solar metallicity is free from degeneracy with mass loss. In fact, in our TP-AGB models $\lambda$ is the main free parameter, while the mass-loss rates for carbon stars do not contain any adjustable efficiency factor. Also, the effect of rotation should not play a substantial role in the IFMR over the $M_i$ range of interest, unlike at higher masses ($M_i > 2.7\ M_\odot$)[34]. Finally, we stress that other models that adopt a much higher $\lambda$ in TP-AGB stars with $M_i \lesssim 2 M_\odot$, or mass-loss rates that do not depend on $C - O$, miss entirely the observed kink (see Extended Data Fig. 2).

Looking at the calibrated 3DU relation (Fig. 4a), we see that $\lambda$ is expected, on average, to increase with the stellar mass for $M_i \lesssim 3 M_\odot$, a trend that is predicted by TP-AGB models in the literature[5]. Our analysis also indicates that this positive correlation is interrupted in two mass intervals. In a narrow range around $M_i \simeq 1.8 - 1.9 M_\odot$, just where the peak of IFMR kink is placed, the data suggest the existence of a local slight minimum in $\lambda$. To interpret this result we need to consider a few key aspects of stellar structure and evolution. The location of such a peak coincides with the initial mass limit, $M_{\mathrm{HeF}}$, that divides the class of low-mass stars (characterised by electron degeneracy in the helium cores left at end of the hydrogen-burning phase and the subsequent He-flash at the tip of the RGB), from the class of intermediate-mass stars (with helium cores that obey the equation of state for the classical ideal gas). Stellar models show that such dichotomy in the equation of state leaves a fingerprint in the mass of the hydrogen-exhausted core, which reaches a minimum value at the stage of central helium ignition for $M_{\mathrm{i}} = M_{\mathrm{HeF}}$. This signature is retained also at later evolutionary stages, up to the onset of the TP-AGB phase. In fact, the core mass at the first TP, $M_{\mathrm{c,1TP}}$, as a function of the initial mass (magenta curve), also exhibits a minimum at $M_{\mathrm{i}} = M_{\mathrm{HeF}}$ [5,35].

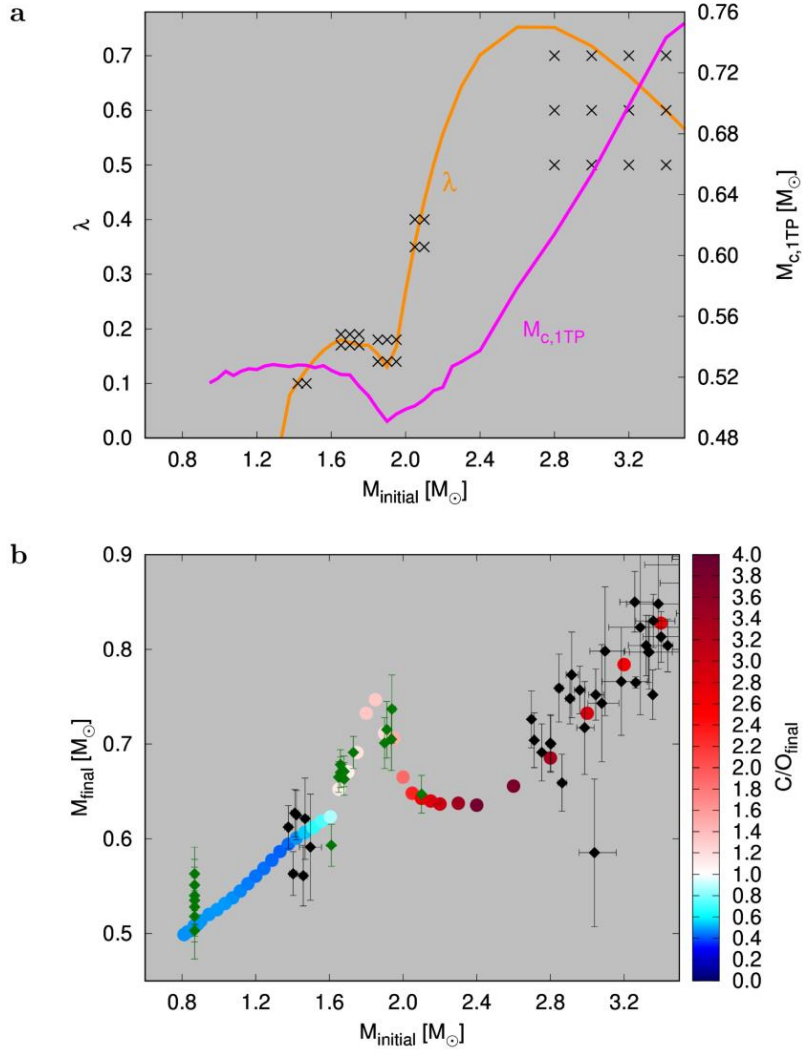

Figure 4: Calibration of the 3DU efficiency and the resulting theoretical IFMR. **a**, The values of the average $\lambda$ (black crosses) that recover the semi-empirical IFMR data and a proposed best-fit curve for $\lambda$ (orange line) as a function of the initial mass; the core mass at the first TP (magenta line). **b**, The theoretical IFMR obtained with the calibrated $\lambda$ relation, colour-coded as a function of the final C/O (right colour bar). The underlying semi-empirical IFMR (diamonds with errors bars) is the same as in Fig. 1, with the 7 newly discovered and 12 newly analysed WDs shown in green. Error bars cover a range of $\pm 1\,\sigma$ (see Table 1).

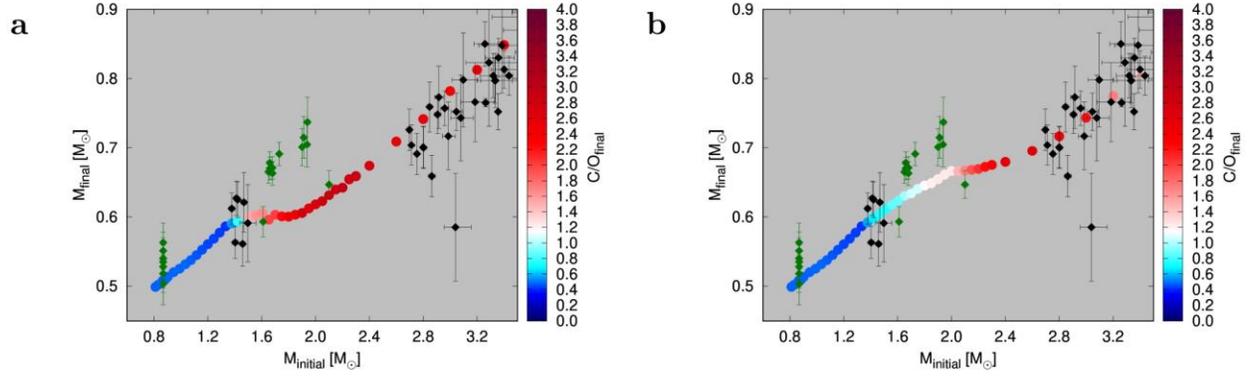

Extended Data Figure 2: Examples of theoretical IFMRs that fail to account for the kink in the semi-empirical IFMR. **a**, Too high efficiency of the 3DU in low-mass stars: $\lambda = 0.5$ is assumed for all models that experience the 3DU. **b**, Mass loss insensitive to the photospheric chemical composition: the B95 mass-loss formula is applied to all models, irrespective of the photospheric C/O. The semi-empirical IFMR is the same as in Fig. 1, with error bars covering a range of $\pm 1\,\sigma$.

From theory, we know that the core mass is a leading parameter of the TP-AGB evolution, and it is expected to affect the 3DU. In particular, lower values of $M_c$ correspond to weaker TPs (measured by the post-flash luminosity peak)[36], hence shallow 3DU events. Consistent with this expectation, our calibration of the 3DU efficiency indicates that the minimum in $\lambda$ occurs close to the minimum in $M_{c,1TP}$. At the same time our evolutionary models predict that such minimum characterises the stellar progenitors of the observed IFMR kink. Putting these two findings together, we propose that the observed peak in WD mass was produced by MW carbon stars with initial masses close to $M_{HeF}$, the transition limit between the classes of low- and intermediate-mass stars. These carbon stars experienced a modest carbon enrichment, an inefficient dust production, mild winds, and a relatively long carbon-star phase.

The positive correlation between $\lambda$ and $M_i$ (Fig. 4a) breaks also at larger stellar masses, $M_i \gtrsim 3 M_\odot$, where $\lambda$ reaches a maximum and then starts to decrease. This latter trend is in line with some existing TP-AGB models[37,38] and earlier studies on the IFMR[8].

## 4  A few remarks about mass loss

Finally, we would like to draw attention to a few aspects about the predicted mass-loss drop (see Fig. 2). First, the amplitude of the drop also depends on the mass-loss rate attained during the phases with $C/O \lesssim 1$, immediately preceding the carbon star formation. If during these stages a dust-driven wind (involving silicate species, for instance) is not yet well developed, the size of the drop may be modest. The results shown here for $C/O < 1$ are based on a widely used, but likely simplistic, mass-loss formula[39]. While important progress has been recently made[40], new detailed stellar wind models for M- and S-type stars (including both SRV- and Mira-like pulsations) are highly desirable for future studies.

Second, it is worth specifying that in our evolutionary calculations we treat the mass-loss stages with $0.85 \lesssim$ C/O $\lesssim 1$ in the same way as those with C/O < 0.85, since suitable wind models for S-type stars are still unavailable. Anyhow, a reasonable guess is that when C/O is quite close to unity while oxygen is still less abundant than carbon, a drop in mass loss may already occur due to the lack of ordinary silicates[24] (see also Fig. 3 for C/O . 1). This circumstance may explain the existence of S-type SRVs with blue colours, $K - [22] < 1$ (see cyan triangles in Supplementary Fig. 6 and related discussion; here [22] is the 22 $\mu$m band of the Wide-field Infrared Survey Explorer (WISE) space observatory).

Lastly and more importantly, the results presented in this work do not depend critically on the size of the mass-loss drop as long as the rates remain sufficiently low (below super-wind values). It makes little difference that the rate is $10^{-10}$ M$_\odot$ yr$^{-1}$ or $10^{-7}$ M$_\odot$ yr$^{-1}$ given that, for typical inter-pulse periods of the order of $10^5$ yr, the reduction of the envelope is small and the evolution proceeds temporarily almost at constant mass. What really matters is the onset of the dust-driven wind in carbon stars, which occurs when $C - O > (C - O)_{min}$ (depending on current values of luminosity ($L$), effective temperature ($T_{eff}$), and $M$). This is mainly controlled by the rapidity with which the atmosphere is enriched in carbon, therefore by the efficiency of the 3DU and the current envelope mass, which strengthens our $\lambda$–calibration.

## 5  Summary and conclusions

A new thorough analysis of a few WDs in old open clusters with turn-off masses over the range from $1.6 M_\odot \lesssim M_i \lesssim 2.1 M_\odot$ has revealed that the IFMR exhibits a non-monotonic component, with a peak of $M_f \approx 0.70 - 0.75 M_\odot$ at $M_i \simeq 1.8 - 1.9 M_\odot$. It happens just in proximity of the transition mass, at $M_i \simeq M_{\mathrm{HeF}}$, predicted by stellar structure models.

The proposed physical interpretation is that the IFMR kink marks the formation of solar metallicity low-mass carbon stars. These latter experienced a shallow 3DU ($\lambda \simeq 0.1-0.2$) during the TP-AGB phase, so that the amount of carbon dust available to trigger a radiation-driven wind was small and $\dot{M}$ remained mostly below a few $10^{-6}\ M_\odot\ \mathrm{yr}^{-1}$. These circumstances led to a prolongation of the TP-AGB phase with the consequence that fairly massive WDs ($M_f > 0.65 M_\odot$), larger than commonly expected, were left at the end of the evolution. The peak of the IFMR kink corresponds to the minimum in the 3DU efficiency ($\lambda \approx 0.1$), hence to the lowest carbon enrichment.

Considering the observed properties of carbon stars in the MW we suggest that the progenitors of the IFMR kink spent a fraction of their carbon star phase as semi-regular variables, characterised by small amplitude pulsations, low C/O, small mass-loss rates and low terminal velocities.

The observed non-linearity of the IFMR at $M_i \simeq 1.8-1.9 M_\odot$ and its interpretation in terms of carbon star formation may have important consequences in the framework of galaxy evolution.

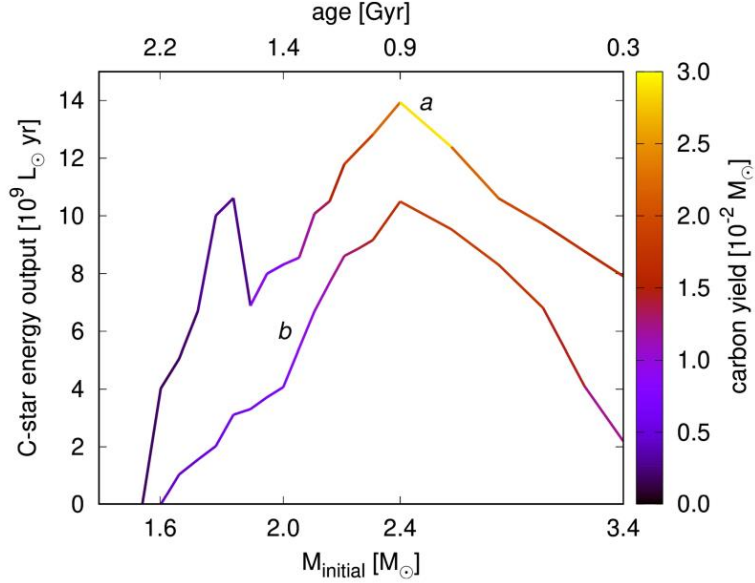

Figure 5: Integrated energy output emitted from the carbon star phase. Predictions are derived by integrating the stellar luminosity along each track over the carbon star lifetime. They are shown as a function of the initial stellar mass (bottom axis), and the corresponding age (top axis). Curve *a* refers to the present calibrated carbon star models (using the $\lambda$-relation shown in Fig. 4a), curve *b* shows the results obtained assuming a mass-loss recipe[39] insensitive to the surface chemical composition. The curves are colour-coded according to the yields of carbon newly synthesised during the TP-AGB phase.

In fact, the IFMR kink happens to occur just around a critical age, $\approx 2$ Gyr, which characterises the oldest stellar populations hosted in galaxies at redshift larger than 3 and where carbon stars may appreciably contribute to the integrated galaxy light[41–43]. Figure 5 shows that the energy output from the carbon star phase at solar metallicity is higher than otherwise predicted by models that use mass-loss relations insensitive to the carbon abundance. In particular, the local peak in WD mass translates into a larger contribution from carbon stars, at ages around $1.5 - 2.0$ Gyr, to the integrated light of galaxies similar to the MW. Dedicated galaxy studies based on population synthesis models are needed to further investigate this aspect.

The impact on the chemical ejecta is also notable. Close to the WD mass peak, the amount of newly synthesised carbon expelled by stars with $1.6 \lesssim M_i/M_\odot \lesssim 1.9$ is quite low due to the little efficiency of the 3DU and the small envelope masses, while it increases by several factors for $M_i > 2 M_\odot$. In conclusion, the progenitors that populate the IFMR kink are expected to be potentially important contributors to the galaxy emitted light and modest sources of carbon (in the form of gas and dust) at the same time.

Finally, we note that the IFMR kink has important consequences for the interpretation of the stellar progenitors of observed WDs in the mass range of $0.65 \lesssim M_f/M_\odot \lesssim 0.75$. Very common in the field and in binary systems, these WDs may have stellar progenitors with two, if not three, possible $M_i$.

**Peer review information** Nature Astronomy thanks Krzysztof Gesicki and the other, anonymous, reviewer(s) for their contribution to the peer review of this work.

**Acknowledgements** P. M., S. B., Y. C., L. G., G. P., M. T., B. A. acknowledge the support from the ERC Consolidator Grant funding scheme (project STARKEY, grant agreement No 615604). P.-E. T has received ERC funding under the European Union's Horizon 2020 research and innovation programme (grant agreement No 677706 - WD3D).

**Author contributions** P. M. designed and performed the theoretical research, ran the TP-AGB models and the population synthesis simulations, and provided the interpretation of the new IFMR data in terms of stellar evolution; J. D. C. performed the Keck observations, processed the data, analysed the cluster parameters, spectroscopically analysed the DAs, and determined memberships; J. L. C. identified the likely WD candidates for observations and assisted with the cluster parameter analysis; J. K. coordinated the observational and theoretical work and provided expertise; P.-E. T. provided the DA WD atmospheric models and fitting program and his expertise; E. R.-R. assisted with Keck observations, P. B. provided the DB WD atmospheric models and fit the DB parameters; S. B. provided expertise and help in implementing the massloss grid of dynamical atmospheres for carbon stars in the `COLIBRI` code; Y. C., A. B., L. G., G. P., M. T. contributed to the development of the stellar models and the discussion of the results. S. C. contributed his WD photometric analysis expertise and his publicly available Python 3 module was used for the photometricbased derivation of WD parameters. B. A. provided expertise and the molecular opacity data to model the atmospheres of carbon stars. P. D. T. implemented the WD models in the populations synthesis simulations.

**Competing Interests** The authors declare that they have no competing financial interests.

**Correspondence** Correspondence and requests for materials should be addressed to P. Marigo (email: paola.marigo@unipd.it).

**Data Availability** The data that support the plots within this paper and other findings of this study are available from the corresponding author upon reasonable request. Montreal WD cooling models are publicly available from the URL http://www.astro.umontreal.ca/ bergeron/CoolingModels/. The pulsation periods are computed with fitting relations based on publicly available models that can be found at the URL http://starkey.astro.unipd.it/pulsation_models.html.

Code Availability The stellar evolution codes `PARSEC` and `COLIBRI` are not publicly available. The mass-loss routine for carbon stars can be found at the URL https://www.astro.uu.se/coolstars/TOOLS/MLRroutines/C/. The code to compute the dust-grain growth in the outflows of AGB stars can be retrieved from the URL http://www.ita.uni-heidelberg.de/ gail/agbdust/agbdust.html. The code used here to calculate photometry-based white dwarf parameters is available from the repository: https://github.com/SihaoCheng/WD_models.

## Methods

### White Dwarf and Cluster Analyses

Gaia DR2[44,45] analysis of the NGC 752 and Ruprecht 147 clusters, with supplements from the deeper Canada-France-Hawaii Telescope photometry of Ruprecht 147[46], have been used to both identify WD candidates and their membership likelihood using proper motions and, for the brighter candidates, their parallaxes. Using two half nights of Keck I/Low Resolution

Imaging Spectrometer (LRIS) analysis on 11 August and 8 September 2018, we observed ten WD candidates in Ruprecht 147 and one candidate in NGC 752. Eight of the ten observed Ruprecht 147 candidates were found to have hydrogen-rich atmospheres (WDs of type DA), one showed a helium-rich atmosphere (WD of type DB), and one is consistent with a background star. Observations of this DB's H$\alpha$ region confirms that there is no hydrogen in its atmosphere. The one NGC 752 candidate is a DA. These WDs have been spectroscopically analysed using the standard DA[47] and updated DB[48,49] methods, and their spectroscopic-based $T_{eff}$, $\log g$ (where $g$ is the surface gravity) and $M_f$ are given in Table 1.

We then applied these spectroscopic parameters to the Montreal cooling models[50] to further test membership and single-star status using the luminosities, colours, and cooling ages. In comparison to the observed photometry, this tells us whether or not these WDs are consistent with being at the same distance, reddening, and extinction as the cluster. In agreement with their proper motions and parallaxes, when available, the one NGC 752 DA and five of the Ruprecht 147 DAs and its one DB are consistent with single-star cluster membership. This DB is the first spectroscopically analysed DB with confirmed membership in a star cluster and it is at remarkable agreement in mass with the five Ruprecht 147 DAs at the same $M_i$. Even though we acknowledge that 3D models should ideally be used in the spectroscopic analysis of DBs at these $T_{eff}$[51], such 3D corrections should be analysed in conjunction with updated van der Waals broadening[49]. However, it is reassuring that the DA and DB masses agree so well because we should expect that at these lower masses the DB IFMR is identical to that of DAs due to their consistent field mass distribution peaks[52].

Lastly, the three remaining observed Ruprecht 147 candidate DAs are at distances comparable to but not in agreement with the cluster. Their observed magnitudes are too faint rather than too bright relative to their spectroscopic-based luminosities, and hence they are also not potential binary cluster members. Moreover, all three have masses of ~0.60 $M_\odot$, inconsistent with the six confirmed members at ~0.67 $M_\odot$ and consistent with the most likely mass of WD-field contaminants. Therefore, we consider them to be non-members.

In Supplementary Fig. 3 we photometrically analyse the cluster turnoffs using Gaia-DR2 based cluster membership and Gaia photometry using our previously adopted cluster analysis techniques[15,53]. For consistent analysis with the presented evolutionary models of carbon star evolution, we derive $M_i$ using both cluster ages and evolutionary timescales based on the PARSEC v1.2S + COLIBRI PR17 isochrones[54,55].

To expand on these spectroscopic techniques, we also take full advantage of the precise photometry, distances, and extinctions available for both these WDs and their clusters. For the Ruprecht 147 and NGC 752 WDs, we adopt the Gaia and CFHT photometry and our fit cluster parameters (Supplementary Fig. 3), and for the more distant NGC 6121 (metal-poor globular cluster M4), NGC 6819, and NGC 7789 WDs we adopt the Johnson BV photometry with our group's previously analysed cluster parameters[15]. When atmospheric composition is known, WD photometric-based parameters are precise with high-quality absolute magnitudes and colours, which are based on each WD's observed magnitudes corrected for cluster distance, extinction, and reddening. However, photometry by itself is not able to determine atmospheric composition (unless UV filters are available), cluster membership, or single-star status. Additionally, our available colours of Johnson B-V or Gaia's BP-RP provide only imprecise $T_{\text{eff}}$ determinations at these high effective temperatures.

To overcome these challenges, here we have continued to use the strengths of spectroscopy to determine atmospheric composition, $T_{\text{eff}}$, and spectroscopic mass, and the combination of these parameters with photometry to test membership and single star status. However, these atmospheric compositions and spectroscopic $T_{\text{eff}}$ are now then also combined with a WD's determined absolute magnitude to provide an independent measurement of WD radius, and hence mass, based on the Montreal cooling models.

This photometric mass is independent of the spectroscopic mass, even though its determination involves spectroscopic $T_{\text{eff}}$, because the spectroscopic $T_{\text{eff}}$ and spectroscopic $\log g$ errors are not correlated. Additionally, while the conversion of $\log g$ to mass is $T_{\text{eff}}$-dependent, it is weakly so at the moderate temperatures of these observed WDs. These photometric $M_{\text{f}}$ determinations also show no signature of a systematic offset between each WD's spectroscopic $M_{\text{f}}$. Therefore, our adopted $M_{\text{f}}$ for each WD is a weighted combination of its photometric and spectroscopic mass (see Table 1 for each WD's absolute magnitude, photometric $M_{\text{f}}$ and weighted-average $M_{\text{f}}$). Furthermore, because these low-mass WDs at the tip of the cooling sequences in these old open clusters are expected to all have consistent mass, it is reassuring that this combination of photometric and spectroscopic information provides a remarkably consistent $M_{\text{f}}$ for each cluster WD, more so than either the photometric or spectroscopic $M_f$ results on their own. In constructing the weighted-average $M_f$, the spectroscopic and photometric estimates, $M_{fS}$ and $M_{fP}$, are give a weight $w$ that anticorrelates with the individual error, namely $w_S = 1/\sigma_S^2$ and $w_P = 1/\sigma_P^2$, respectively. The final error $\sigma$ associated to each weighted-average $M_f$ is obtained via the standard relation $\sigma = 1/\sqrt{\sigma_S^{-2} + \sigma_P^{-2}}$. As a consequence, the uncertainties of $M_f$ are reduced. In particular, the scatter in $M_f$ among the WDs belonging to NGC 7789, which populate the region close to the kink peak, shrinks considerably compared to previous work[15]. Lastly, Table 1 gives the cooling age and $M_i$ resulting from this weighted-average $M_f$.

These new semi-empirical IFMR data (see Tab. 1) have now been combined with WDs from our previous publication[15], which have also been more precisely analysed with the addition of their photometry. In addition, the WDs from M67[16,17] (with their $M_i$ adjusted for consistency with our own PARSEC model fits of cluster age and evolutionary timescales) have been included. However, due to limitations with the available M67 WD photometry, only their spectroscopic WD parameters are used.

These low-mass WD data can be described with the fitting relations ($M_i$ and $M_f$ are in units of $M_\odot$):

$$M_f = \begin{cases} 0.447 + 0.103\, M_i & \text{if } 0.85 \leq M_i \leq 1.51 \\ 0.001 + 0.399\, M_i & \text{if } 1.51 < M_i \leq 1.845 \\ 1.367 - 0.342\, M_i & \text{if } 1.845 < M_i \leq 2.21 \\ 0.210 + 0.181\, M_i & \text{if } 2.21 < M_i \leq 3.65 \end{cases}$$

See our previous publications[15,34] for the IFMR at masses of $M_i$ > 3.65 $M_\odot$.

To establish the statistical significance of this kink, as illustrated in Fig. 1 and discussed here, we also rule out simpler IFMR fits of these low-mass IFMR data. We first test that the WD $M_f$ errors are robust and are not underestimates. This is done by comparing the $M_f$ errors to the intracluster $M_f$ scatters, where at these lowest masses these WDs observed at the top of each cluster's cooling sequence can be assumed to have the same true $M_f$ and $M_i$. For NGC 6121, M67, Ruprecht 147, and NGC 7789, in each cluster 57% to 100% of the observed WD $M_f$ are within their individual 1$\sigma$ errors of their cluster weighted mean $M_f$. For all clusters combined, 75% (18 out of 24) of the WDs are within 1$\sigma$ of their cluster's $M_f$ weighted mean. Normal error distributions are that ~68% of data should be within 1$\sigma$ of their distribution's mean. Therefore, the $M_f$ errors are reliable and reproduce well the observed scatter.

Now that we have illustrated the robustness of the $M_f$ errors, these errors act both as a weighting for the IFMR fits of these low-mass data and can illustrate how well various IFMR-fit shapes represent these data. A weighted linear fit of the low-mass data from $M_i$ of 0.85 to 2.85 $M_\odot$ (purple line) creates a poor representation of the data where only 8 of the 26 (31%) low-mass WDs are within their 1$\sigma$ errors of this linear fit, where again we should expect ~68% to be for a representative fit. This fit also only gives a reduced chi-squared of 2.57, representative of underfitting the data.

The next level of complexity for fitting these data is a weighted linear fit from 0.85 to 2.15 $M_\odot$ (magenta line), or comparably from 0.85 to 2.00 $M_\odot$ (excluding NGC 752-WD01; blue line). However, like with our adoption of a kinked fit, both of these cases must still then be followed by a temporarily decreasing IFMR to match back with the Hyades WDs at $M_i$ of 2.7 $M_\odot$ and to not overproduce 0.7 $M_\odot$ WDs in the field. This latter point is illustrated in the Supplementary Section "The WD mass distribution" and Supplementary Figs. 1 and 2. With both fits, 19 of the 26 (73%) and 19 out of 25 (76%) low-mass WDs are now within their 1$\sigma$ errors of these fits, but they are still unlikely representations of the data because these fits overestimate the mass of nearly every M67, NGC 6819, and NGC 752 WD while they underestimate the mass of nearly every Ruprecht 147 and NGC 7789 WD. Such clumpy (non-random) residuals are extremely unlikely for a fit that well represents the data. For reference, the reduced chi-squared values of these two fits are 1.88 and 1.09, respectively, the latter of which does not rule out its goodness of fit, but again the clumpy residuals argue against this being representative of the data.

Finally, the low-mass IFMR ($M_i \leq 2.21 M_\odot$) is described with the 3-piece kink fit given above (black line), which connects to the intermediate-mass $2.21 < M_i \leq 3.65$ and higher-mass

IFMR discussed in greater detail in our previous work[15,34]. Additionally, where the data remain limited between $M_i$ of $2.0 M_\odot$ and $2.7 M_\odot$, the fit is partially guided by the model in Fig. 4b. This 3-piece kink is the simplest fit that can both match the current IFMR data and produce appropriately random residuals (20 of the 26 data WDs (77%) being within $1\sigma$ of the fit and with a reassuring reduced chi squared of 1.02). At the same time, the kink IFMR fit proves to be consistent with the field WD mass distribution in the MW (see Supplementary Fig. 2e). Clearly, further IFMR data for $2.0 \lesssim M/M_\odot \lesssim 2.7$ will be valuable, but the existence of an IFMR kink is becoming established.

**Stellar evolution models**

We computed a set of TP-AGB models with the `COLIBRI` code[35], for 42 selected values of the initial mass in the range $0.8 \lesssim M_i/M_\odot \leq 3.4$, assuming a solar-like initial metallicity[56], $Z = 0.014$. We recall that, according to a standard terminology, $Z$ denotes the total abundance (in mass fraction) of all elements heavier than helium. The initial conditions at the first TP are taken from a grid of `PARSEC` tracks[54]. A key aspect of the TP-AGB models is the integration of the `ÆSOPUS` code[23] inside `COLIBRI`. This allows to calculate on-the-fly both the equation of state for about 800 atomic and molecular species and the Rosseland mean gas opacities (for temperatures $2,500 K \leq T \leq 15,000$ K). In this way, the effects of abundance changes, due to mixing events, on the atmospheric structure can be treated in detail.

On the red giant branch mass loss is computed using the Reimers law[57], with an efficiency parameter $\eta_R = 0.2$. Later, the AGB mass loss by stellar winds is described as a two-stage process[58]. During the early phases, for luminosities below the tip of the red giant branch ($\log L/L_\odot \simeq 3.4$–$3.5$), we assume that mass loss is driven by Alfvén waves in cool and extended chromospheres[59]. Later, as long-period variability develops and the star becomes luminous and cool enough, powerful winds are accelerated by radiation pressure on dust grains that condense in the outer layers of the pulsating atmospheres. In the dust-driven regime the mass-loss rate is computed with different prescriptions according to the surface C/O ratio, namely: a widely-used relation[39] (with an efficiency parameter $\eta_B = 0.01$; B95 in the text) based on dynamical calculations of the atmospheres of Mira-like stars for the O-rich stages when C/O < 1, and a routine based on state-of-the-art dynamical atmosphere models[21,22] for carbon stars when C/O > 1 (also referred to as CDYN in the text). These models, in particular, predict that dust-driven winds in carbon stars are activated only when i) the amount of free carbon, C − O, exceeds a threshold $(C - O)_{min}$, and ii) favourable conditions exist in the extended atmospheres for condensation of dust grains. Inside the grid of wind models for carbon stars adopted here, the threshold in carbon excess is predicted to vary within a range, $8.2 \lesssim (C - O)_{min} \lesssim 9.1$, the exact value depending on other stellar parameters, namely mass, luminosity, and effective temperature. At solar-like metallicity, $Z = 0.014$, assuming that oxygen is not altered by the 3DU, such a range in $(C - O)_{min}$ translates into a range of the minimum C/O required for radiation-driven winds, $1.3 \lesssim (C/O)_{min} \lesssim 2.7$.

As long as suitable conditions for the activation of the dust-driven wind are not fulfilled, typically during the early stages of carbon stars (characterised by low C − O, low $L/M$ ratios, and relatively high $T_{\rm eff}$) the CDYN prescription cannot be applied. In these cases we reasonably assume that stellar winds are sustained by pulsations alone[33]. Pulsation-driven mass loss is described by fitting a set of dynamical models for dust-free atmospheres[60], and expressing the mass-loss rate [$M_\odot$ yr$^{-1}$] with the form[61] $\dot{M} = \exp(a\,M^b R^c)$, where $M$ and $R$ denote the star's mass and radius in solar units ($a = -789$, $b = 0.558$, and $c = -0.676$). The resulting rates are typically low, from $\approx 10^{-9} - 10^{-7}\,M_\odot$ yr$^{-1}$, but values of $\approx 10^{-6}\,M_\odot$ yr$^{-1}$ can be reached for suitable combinations of stellar mass, luminosity and effective temperature.

The set of mass-loss prescriptions described above constitutes our standard choice for the models presented in this work. For the purpose of discussion, we also apply to carbon stars additional options that do not depend on the photospheric C/O, namely: the B95 formula or the semi-empirical formalism[62] (VW93 in the text) that relates the mass-loss rate to the fundamental mode period of pulsating AGB stars.

The 3DU is described with a parametric approach[35] that determines the onset and the quenching of the mixing events, and their efficiency $\lambda$. A 3DU episode takes place when a temperature criterion is met, i.e., provided the temperature at the base of the convective envelope exceeds a minimum value, $T_{\rm b}^{\rm dred}$, at the stage of the post-flash luminosity maximum. We adopt[35] $\log T_{\rm b}^{\rm dred} = 6.4$.

To perform a systematic exploration of the effect of the 3DU efficiency on the IFMR, similarly to the approach introduced in an earlier study[8], we ran a large grid of TP-AGB models varying $\lambda$ from 0.05 to 0.7 in steps of 0.05 or 0.1, for all values of $M_i$ under consideration. To avoid additional free parameters in the description of the 3DU, we make the simple assumption that the efficiency $\lambda$ is constant during the TP-AGB phase as long as the temperature criterion with $T_b^{\mathrm{dred}}$ is fulfilled. Therefore, the values of $\lambda$ should be considered as a measure of the average efficiency of the 3DU experienced by a star of given initial mass.

The results are shown in Extended Data Fig. 1. We see that a sizeable scatter is produced, as a consequence of the changes in both core mass and surface chemical composition. The case $\lambda = 0.05$ does not form any carbon star at any $M_i$. As to the models with final C/O > 1 we see that, on average, increasing $\lambda$ results in less massive WDs at a given $M_i$. The reason is twofold.

On one side, a more efficient 3DU reduces the net core-mass growth (by an amount $\lambda \times \Delta M_c$ at each mixing event). On the other side, it enriches the atmosphere with more carbon, which favours the onset of the dust-driven wind with consequent earlier termination of the TP-AGB phase. The present model grid is used to calibrate $\lambda$ as a function of $M_i$, as discussed in the main text.

# Supplementary Information

## Carbon star formation as seen through the non-monotonic initial-final mass relation

*Paola Marigo, Jeffrey D. Cummings, Jason Lee Curtis, Jason Kalirai, Yang Chen, Pier-Emmanuel Tremblay, Enrico Ramirez-Ruiz, Pierre Bergeron, Sara Bladh, Alessandro Bressan, Léo Girardi, Giada Pastorelli, Michele Trabucchi, Sihao Cheng, Bernhard Aringer, Piero Dal Tio*

## The WD mass distribution

The observed field-WD mass distribution (WDMD) in the MW can be employed to further check the impact of the IFMR kink and, more generally, of any assumed IFMR. Ideally, one may simultaneously use the observed IFMR data and the WDMD to constrain the coefficients of a parametric functional form expressing $M_f$ as a function of $M_i$. However, at present the application of this method is made troublesome by the fact that a detailed consideration of the WDMD would introduce additional parameters to be examined, affecting both the population synthesis simulations (initial mass function, star formation history, chemical enrichment law) and the underlying stellar evolutionary models, in particular the metallicity dependence of all relevant processes involved (3DU and mass loss). This approach would introduce further uncertainties in our analysis with the consequence that the significance of our main observable, the IFMR, would be weakened. Therefore, at this stage, our strategy is to adopt the IFMR as the primary calibrator and employ the WDMD as an auxiliary verification tool to test the impact of different fitting forms. The fits examined here are illustrated in Supplementary Fig. 1. They adopt different assumptions concerning the IFMR kink and the behaviour of the IFMR through the region $(2.0 \lesssim M_i/M_\odot \lesssim 2.8)$ that shows a substantial lack of data.

Supplementary Fig. 2 shows the resulting WDMDs for the five considered IFMR fits. All simulations were carried out with the TRILEGAL code,[1] assuming a Kroupa initial mass function,[2] an age-metallicity relation and a star formation history suitable for the MW disk. The stellar evolutionary tracks are taken from the PARSEC[3] and COLIBRI[4,5] databases, tailored to a grid of H-burning Post-AGB tracks and WD cooling sequences.[6] Here we examine the distribution of WDs with masses from $0.55\,M_\odot$ to $1.2\,M_\odot$. We exclude WDs with $M_f < 0.55\,M_\odot$ because their proper consideration would require taking care of various aspects that are still quite uncertain or go beyond the scope of this study, like the metallicity dependence of the IFMR (the metal-poor regime, in particular) and the effects of binary interactions. Within the considered mass range, the SDSS data[7] has 4597 WDs of type DA with a signal-to-noise ratio $> 10$. The simulated distributions were re-scaled to the same number of WDs as observed.

The simplest case (fit #1) corresponds to a monotonic relation that runs through the data over the range $0.8 \lesssim M_i/M_\odot \lesssim 3.1$. The monotonic representation has been the standard choice of all IFMR studies so far. This fit produces a rough approximation of the observations, but results in a WDMD peak located at too high mass, from $0.60\,M_\odot$ to $0.65\,M_\odot$ (Supplementary Fig. 2a). The shift is the consequence of the higher intercept we derive when attempting to describe the observed data through a single linear representation. This inconsistency, in addition to its poor representation of the IFMR data around $1.7\,M_\odot - 2.0\,M_\odot$ further argues against a simple monotonic fit.

As next step we relax the monotonic assumption. We first fit the rising part of the IFMR up to a peak of $M_f \simeq 0.71\,M_\odot$ located at $M_i \simeq 1.8\,M_\odot$, and then, beyond it at larger $M_i$, we test a few trends. We start by adding a plateau (fit #2) of $M_f \simeq 0.71\,M_\odot$ to join the increasing branch starting at $M_i \simeq 2.8\,M_\odot$. In this way we exclude the WD belonging to NGC 752 from the fit. Clearly, the case #2 provides a very poor match of the whole observed WDMD from $M_f \simeq 0.6\,M_\odot$ to $M_f \simeq 0.8\,M_\odot$, with a dramatic excess of WDs of $M_f \approx 0.70 - 0.75\,M_\odot$ (Supplementary Fig. 2b).

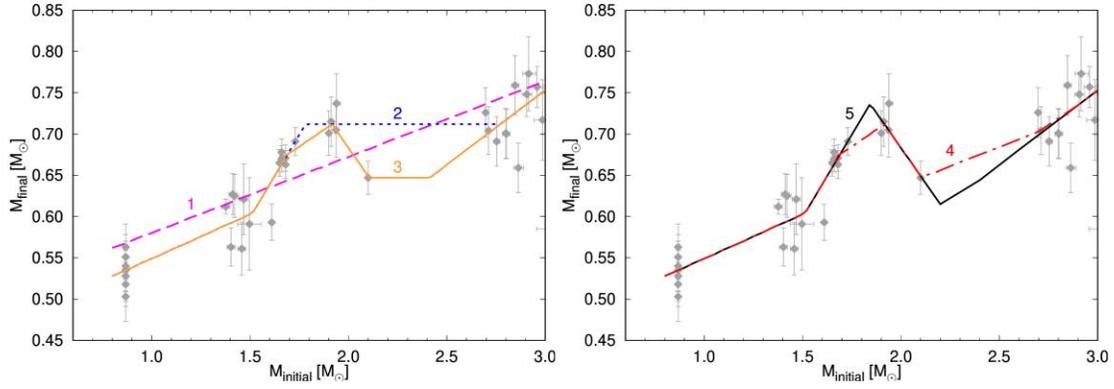

**Supplementary Figure 1:** Sample of five fitting relations compatible with the semi-empirical IFMR, which mainly differ in the description of the interval $1.7 \lesssim M_i/M_\odot \lesssim 2.8$. They are meant to explore the effect of introducing a non-monotonic feature in the IFMR, and to analyse its impact on the WDMD when varying the width and the amplitude of the associated kink. The error bars in the semi-empirical IFMR cover the range of $\pm 1\sigma$.

Recognised that the plateau with $M_f \simeq 0.71\,M_\odot$ over the range $1.8 \lesssim M_i/M_\odot \lesssim 2.8$ does not work, we reintroduce the NGC 752 data and impose that the fit #3 goes through the weighted averages of the WDs in the clusters over the interval $1.65 \lesssim M_i/M_\odot \lesssim 2.1$. After reaching a peak of $M_f \simeq 0.71\,M_\odot$ at $M_i \simeq 1.91\,M_\odot$, the fit #3 decreases down to recover the mass $M_f \simeq 0.65\,M_\odot$ of N752-WD01 at $M_i \simeq 2.1\,M_\odot$. From this point a small plateau with $M_f \simeq 0.65\,M_\odot$ extends to $M_i \simeq 2.4\,M_\odot$, where it intersects the extrapolated linear relation that describes the rising branch at larger masses ($M_i > 2.8\,M_\odot$). Introducing this kink certainly improves the representation of the IFMR over the relevant $M_i$ range, and at the same time it reduces the discrepancy in the WDMD that affects the previous plateau case (fit #2). We see that also the fit #3 yields an excess of WD with $M_f \simeq 0.65 - 0.70\,M_\odot$, though at a lower level (Supplementary Fig. 2c).

In the previous case (fit #3), we have tested the assumption that the IFMR stays flat over the range between $M_i \simeq 2.1\,M_\odot$ and $M_i \simeq 2.4\,M_\odot$, just where the IFMR is essentially unconstrained due to the lack of data. Therefore, we are driven to analyse other trends in this region. Now we examine the case (fit #4) in which, after reaching a minimum of $M_f \simeq 0.65\,M_\odot$ in correspondence to N752-WD01, the IFMR regains a positive slope connecting to the branch starting at $M_i \simeq 2.8\,M_\odot$. While the performance of the fit #4 is the same as the previous #3 – the two only differ in the region void of data–, the impact on the WDMD is sizeable. With the fit #4 the excess of WDs in the range $M_f \simeq 0.65 - 0.70\,M_\odot$ appears now amplified (Supplementary Fig. 2d).

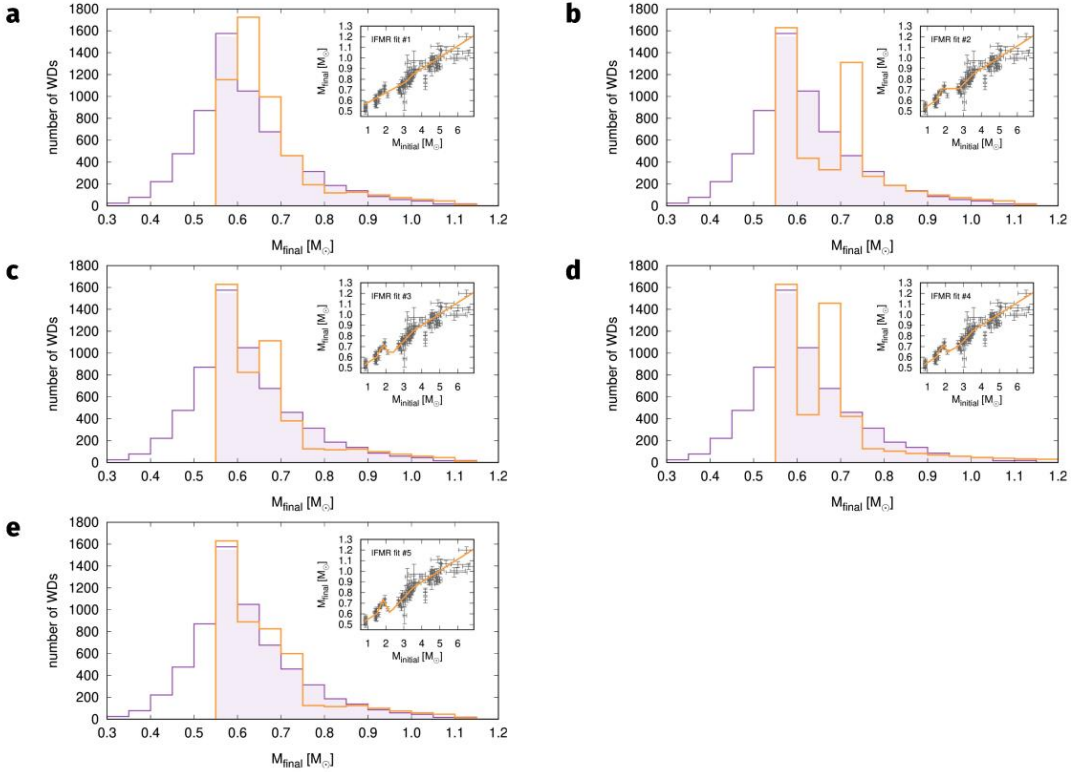

**Supplementary Figure 2:** The field white dwarf mass distribution in the MW. The observed distribution (purple histogram) is taken from the SDSS DR7.[7] Over the range $0.55 \lesssim M_i/M_\odot \lesssim 1.2$ we overplot the results of population synthesis simulations (orange histograms) obtained assuming the same numbers of WDs as observed, and five different IFMR fits, shown in the insets and in Supplementary Fig. 1.

This fact leads us to increase the depth of the kink. With the fit #5 we assume that, after matching N752-WD01, the decrease continues down to a slightly deeper minimum ($M_f \simeq 0.62\,M_\odot$ at $M_i \simeq 2.2\,M_\odot$), beyond which a linear positive trend is re-established so as to pass through the data at intermediate $M_i$. In this way the surplus of WDs with $M_f \simeq 0.7\,M_\odot$ is reduced to a satisfactory level. The resulting theoretical distribution much better matches the observed peak and approximately recovers the intermediate to high-mass wing (Supplementary Fig. 2e).

From the tests performed we conclude that a reasonably good reproduction of both the IFMR and the WDMD supports the existence of a kink in the IFMR. The decline that follows the peak at $M_i \simeq 1.8 - 1.9\,M_\odot$ is needed to avoid the occurrence of a sizeable excess of WDs with $M_f \simeq 0.7\,M_\odot$. Clearly, a more precise assessment of the extension (in $M_i$) and the amplitude (in $M_f$) of the kink will require further efforts involving additional observational data across the range $2.1 \lesssim M_i/M_\odot \lesssim 2.8$, coupled to a more detailed modelling of the WDMD. Future studies will be devoted to address these aspects.

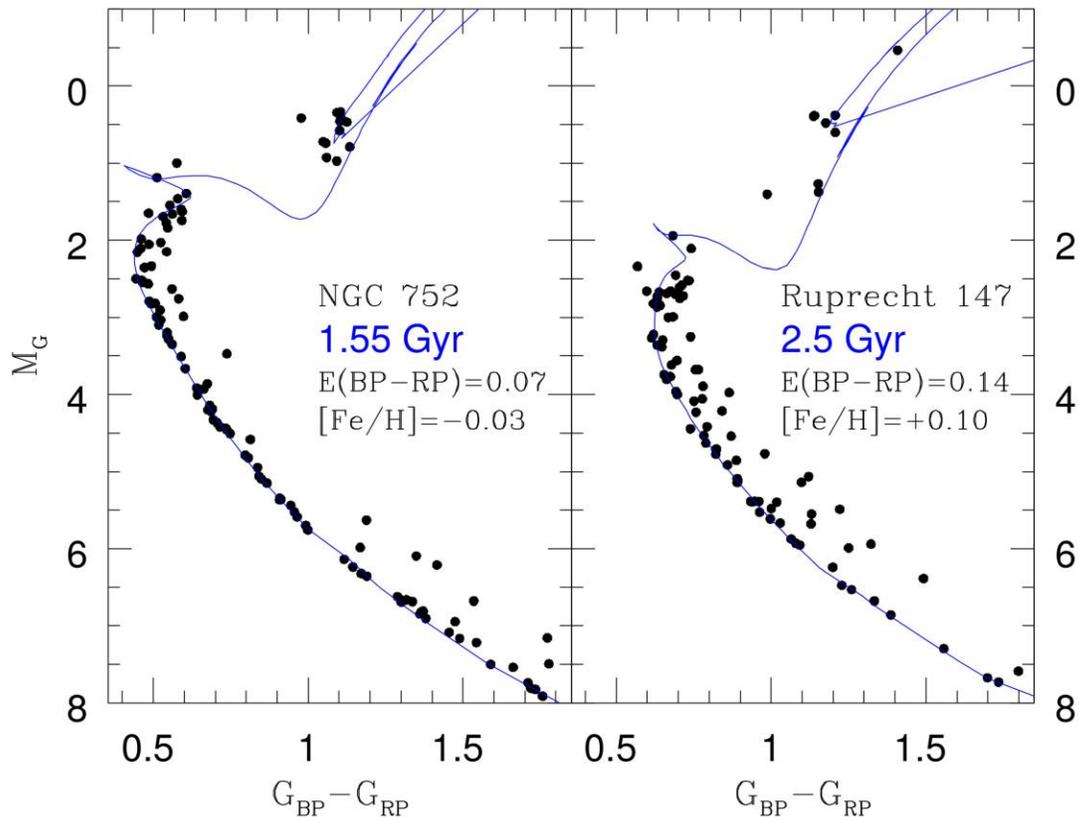

**Supplementary Figure 3:** Colour-magnitude diagrams of the two old open clusters hosting the 7 newly discovered WDs. The cluster ages and reddenings of NGC 752 and Ruprecht 147 are fit using Gaia DR2 photometry, where only likely cluster members are displayed. The fit isochrones are the PARSEC v1.2S + COLIBRI PR17 isochrones[3,8] with the adoption of the revised Gaia DR2 passbands.[9] The cluster metallicities are derived from spectroscopic analyses in the literature,[10–13] which reassuringly match well the main sequence trends and colours of the giants.

# Other supporting evidence: Galactic semi-regular variables

Observed properties of Galactic AGB stars belonging to the class of semi-regular variables provide hints that support the proposed physical explanation. Let us examine the most relevant facts.

**Progenitors of the IFMR kink.** The picture proposed to explain the existence of the IFMR kink relies on the existence of carbon stars with small carbon enrichment and mild mass loss. In fact, Galactic carbon stars with low $C - O$ ($< 8.2$) and low C/O ($< 1.5$) are observed,[14] and have spectra dominated by dust-free stellar photospheres[15] (see also Fig. 3, main article). They have moderate mass-loss rates[16] (median value of $\simeq 2 \times 10^{-7}\,M_\odot\,yr^{-1}$), not exceeding $10^{-6}\,M_\odot\,yr^{-1}$, as well as low expansion velocities,[16] typically with $v_{\mathrm{exp}} < 10\,\mathrm{km\,s^{-1}}$ (median value $\simeq 7 - 8\,\mathrm{km\,s^{-1}}$). They are long-period variables,[14,16] with pulsation periods $P < 400$ days (median value $\simeq 200$ days), mostly belonging to the classes of Irregulars (Lb) and Semi-regulars (Sra, Srb).

Overall, these facts converge to the conclusion that, as long as C/O remains low enough, carbon stars have weak winds, not sustained by radiation pressure on dust grains, but likely, by low-amplitude stellar pulsation.[17–19]

All the above-mentioned properties belong to the group of Galactic carbon stars inside the light blue region *B* of Supplementary Fig. 4 and, according to the proposed scenario, should also characterise the carbon-rich progenitors of the WDs we see now populating the IFMR kink. To better investigate this point we computed the evolution of the terminal wind velocity along a few selected TP-AGB tracks by means of a stationary AGB wind model,[20] which follows the growth of the dust grains as a function of stellar parameters (C/O, $M$, $L$, $T_{\mathrm{eff}}$, $\dot{M}$). The adopted density profile across the circumstellar envelope was modified to include the effects of shocks in the inner wind.[21] We see that the model with $M_{\mathrm{i}} = 1.85\,M_\odot$ spends most of its carbon-star phase inside or close to the region *B*, never exceeding $\dot{M} \simeq$ few $10^{-6}\,M_\odot\,yr^{-1}$ and $v_{\mathrm{exp}} \simeq 10\,\mathrm{km\,s^{-1}}$. The predicted low values of the carbon excess are in agreement with the measured chemical abundances of this group of carbon stars.

These results support the proposed picture: the observed mild carbon enrichment and moderate mass loss, which define this sub-sample of SRV stars, set the suitable conditions for a prolongation of the carbon-star phase, hence for the growth of the core mass. We note that stellar models with larger initial masses ($M_{\mathrm{i}} = 2.0\,M_\odot$ and $M_{\mathrm{i}} = 2.8\,M_\odot$) may also cross the region *B*, but they soon leave it and attain larger carbon excess, as well as higher mass-loss rates and outflow velocities. Their properties are consistent with the group of infrared carbon stars, dominated by the presence of dust in their observed spectra,[22] such as the well-known IRC+10126[23] which is characterised by C/O $\approx 2.5$, as inferred from molecular emission lines.[24]

**Mass loss of Galactic carbon stars.** A closer look at the infrared properties of long-period variables in the Milky Way provides further support to the proposed mass-loss scenario for carbon stars.

To this aim it is useful to consider the colour $K-[22]$, where $[22]$ is the $22\,\mu$m band of the Wide-field Infrared Survey Explorer (WISE) space observatory.[25] As pointed out in a recent study,[26] the colour $K-[22]$ is an indicator of the mass-loss rate, as can be appreciated in Supplementary Fig. 6, which combines photometric data with recent mass-loss rate measurements, based on molecular line emission, for a group of MW long-period variables. We see a clear positive correlation between $K-[22]$ and $\dot{M}$. The SRVs mostly populate the branch steeply rising in mass-loss rate ($\dot{M}$ up to $\approx 10^{-6}\,M_\odot\,yr^{-1}$) characterised by low values of the infrared colour, up to $K-[22] \approx 3$. Then, the positive trend extends with the Mira stars which draw an elongated branch at larger mass-loss rates, reaching quite red colours, up to $K-[22] \approx 11$.

We fit the data of Supplementary Fig. 5 by adopting a functional form $y = a/(1 + b \times x) + c$, where $x = K-[22]$, $y = \log(\dot{M}_6)$ and $\dot{M}_6$ is the mass-loss rate in units of $10^{-6}\,M_\odot\,yr^{-1}$. The coefficients of the best-fitting solutions are $a = -13.068$, $b = 0.814$, and $c = 3.205$ for M-type stars and $a = -4.036$, $b = 0.273$, and $c = 1.554$ for C-type stars. We use the two relations to colour a few evolutionary tracks and compare the results with observations of AGB variable stars in the diagram that relates the pulsation period with the infrared colour $K-[22]$ (Supplementary Fig. 6). Predicted periods for different radial models are obtained on the basis of an extensive theoretical work.[35] We apply the predicted fundamental period, $P_0$, an empirical correction ($\log(P_0^{cor}) = \log(P_0) + 2.341 \times \log(T_{\text{eff}}) - 8.33$), which has shown to yield a satisfactory reproduction of the sequence C, mostly populated by Miras, in the period-luminosity diagram of the Small Magellanic Cloud.[36] In this respect it is worth underlining that the main focus here are the not the Mira variables but rather the SRVs, as is clear from the previous analyses on dust condensation (Fig. 3, main article), terminal wind velocities (Supplementary Fig. 4), as well as from the discussion that follows.

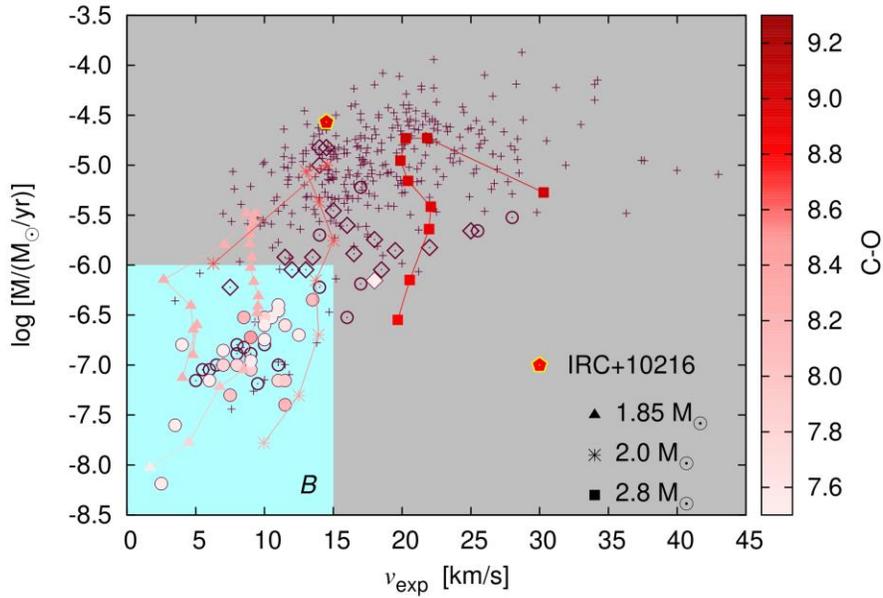

**Supplementary Figure 4:** Mass-loss rates versus terminal wind velocities of Galactic intrinsic carbon stars. The observed data include a sample[16] of optically-bright carbon stars (circles: Semi-regular and Irregular variables; diamonds: Mira variables) and an IR-complete sample[22] (crosses). When the photospheric abundances of carbon and oxygen are available from spectroscopic analyses,[14,27] the stars are colour-coded according to the carbon excess, $C - O$. The light-blue region (labelled with *B*) include Galactic carbon stars with low C/O and $C - O$, moderate $\dot{M}$ and low $v_{exp}$, small-amplitude pulsation, mostly dust-free spectra. The median value of the measured carbon excess is $\simeq 7.5$, well below the threshold of dust-driven winds. Carbon stars outside the *B* region, such as the infrared carbon star IRC+10126 (pentagon), exhibit larger values of all these parameters (C/O, $C - O$, $\dot{M}$, $v_{exp}$, $P$), and their spectra are dominated by dust. For comparison, we overplot three evolutionary tracks during the carbon-star phase, with initial masses (in $M_\odot$) as indicated. The track with $M_i = 1.85\,M_\odot$ and initial solar metallicity runs mostly within this region and exemplifies the evolution of the progenitor carbon stars that populate the IFMR kink with their WD remnants.

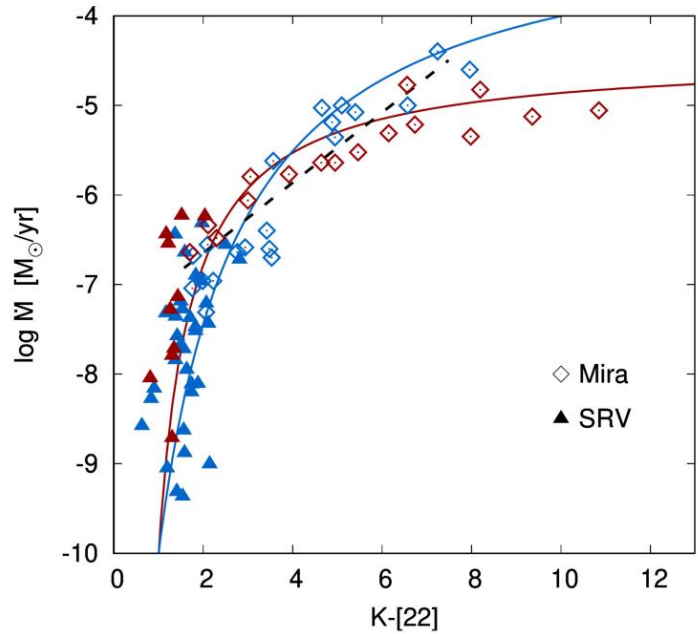

**Supplementary Figure 5:** Relation between the gas mass-loss rate and the infrared colour K-[22] for a sample of MW semi-regular (triangles) and Mira variables (squares). Mass-loss measurements for Miras are taken from an extensive compilation of data,[26] but restricted to the most recent analyses.[28–32] Data for SRVs are obtained from a new study[33] that employs DR2 Gaia distances.[34] Stars with spectral types M and C are coloured in blue and red, respectively. The solid lines are two fit relations with the same colour coding as for the data, the dashed line represents a linear fit (over the approximate range $1.5 \leq K - [22] \leq 7.5$) proposed in a recent work.[26]

The diagram in Supplementary Fig. 6 is helpful to test the plausibility of the proposed mass-loss scenario, including the drop in mass loss during the very early stages of the C-star phase. The choice of the two evolutionary models with initial masses, $1.95\,M_\odot$ and $2.40\,M_\odot$, is motivated by the fact that the former should represent a stellar progenitor of the IFMR kink, while the latter exemplifies the case of a star which gets more efficiently enriched in carbon and populates the IFMR beyond the kink. We also note that the density of points along the tracks is controlled by the sampling of the pulse-cycle phase and it is not indicative of the observational probability, for which dedicated population synthesis simulations are necessary. In addition to our reference CDYN mass-loss prescription (Supplementary Figs. 6a-6b), we also consider TP-AGB models computed with the VW93 and B95 formulas, which do not depend explicitly on the carbon abundance (Supplementary Figs. 6c-6f).

Independently of the assumed mass loss, both TP-AGB tracks with $M_i = 1.95\,M_\odot$ and $M_i = 2.4\,M_\odot$ show a similar evolution of their pulsational properties. During the oxygen-rich stages (with C/O $< 1$) pulsation takes place in the 2$^{nd}$ and 1$^{st}$ overtone modes with periods $\lesssim 200$ days.

Soon after the transition to the C-star domain (with C/O $\gtrsim 1$), the evolution is initially still dominated by the 1$^{st}$ overtone modes, until when the fundamental mode becomes excited, remaining dominant up to the end of the quiescent TP-AGB evolution. Longer periods (typically $P_0 > 250$ days) are attained, the colour $K-[22]$ becomes redder and the models cross the sequence populated by the Miras. During these stage transitions back to higher overtone modes may take place along the same pulse cycle, typically when a star is in the post-flash low-luminosity dip.[35]

While this picture holds, in general terms, for all the models shown in Supplementary Fig. 6, important differences emerge in the infrared colours $K-[22]$. During the C/O $< 1$ stages the models (purple) go through the region occupied by the M-type SRVs (blue ellipse). Their colour $K-[22]$ remains rather low, indication that the mass-loss rates are mostly moderate. Relatively red colours ($K-[22] \simeq 3$) are attained only by the $M_i = 1.95\,M_\odot$ at the tip of its 1$^{st}$ overtone sequence, just before the transition to the carbon-star regime.

When the stars become carbon rich, the behaviour of the models radically differ depending on the assumed mass-loss scheme. Models that adopt the CDYN prescription (Supplementary Figs. 6a-6b) experience a drop in mass loss due to the low carbon excess and experience a temporary pulsation-driven moderate wind phase. The mass-loss drop translates into a colour jump: the model with $M_i = 1.95\,M_\odot$ suddenly falls from $K-[22] \approx 3$ at the tip of the 1$^{st}$ overtone O-rich sequence down to $K-[22] \approx 1$-$2$, values that characterise the group of C-type SRVs (red ellipse). A similar behaviour, but with a less extended colour jump, is exhibited by the model with $M_i = 2.40\,M_\odot$.

In summary, the CDYN models are able to reach the region of C-type SRVs as a consequence of the drop in mass loss predicted after the transition to the carbon-star phase. Later, as more TPs take place, the fundamental mode becomes dominant and the models attain larger mass-loss rates evolving through the branch mostly occupied by Mira stars at increasing period and redder $K-[22]$. We also note that the bottom of the fundamental-mode sequence is partly populated by SRVs.

Conversely, the VW93 and B95 models (Supplementary Figs. 6c-6f) do not recover the observed bluer $K-[22]$ colours characteristic of the C-type SRVs and remain outside the red ellipse. This implies that both prescriptions tend to overestimate the mass-loss rates during the early stages of the carbon-star evolution.

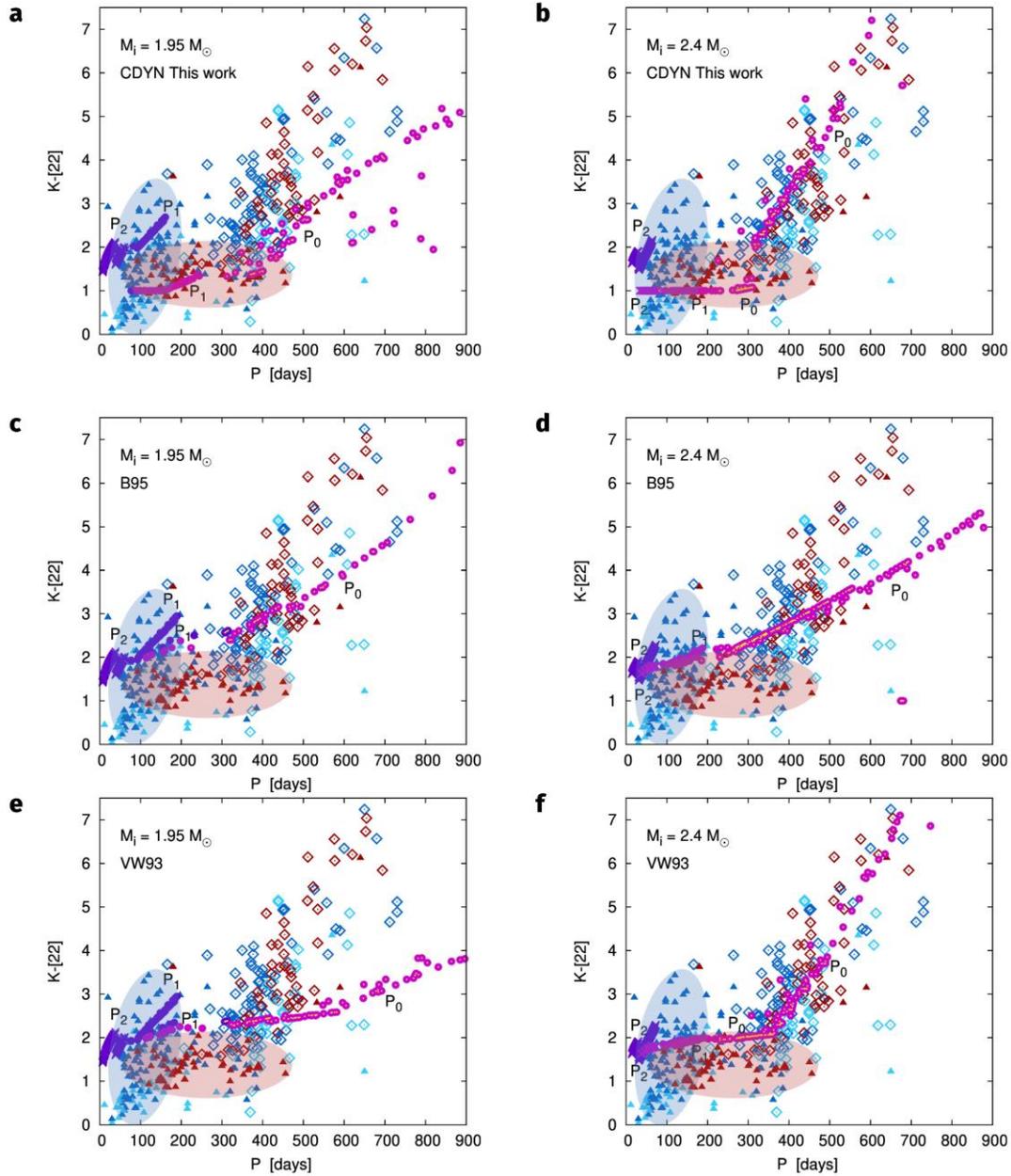

**Supplementary Figure 6:** Infrared colour $K - [22]$ vs pulsation period of AGB stars in the MW. Symbols denote different classes of variables (SRV: filled triangles; Mira: empty squares) while colours correspond to different spectral types (blue: M-type, cyan: S-type, red: C-type). The location of of the bulk of M-type and C-type SRVs is highlighted with two elliptical shaded areas to help the comparison with the models. Here we include the results of TP-AGB evolutionary calculations for models with $M_i = 1.95\,M_\odot, 2.40 M_\odot$. Three descriptions of mass loss are adopted, namely the reference CDYN (**a-b**), B95 (**c-d**) and VW93 (**e-f**). The model results are shown for discrete values of the pulse-cycle phase $\phi$ (for each TP $\phi$ goes from $0.1$ to $1$ in steps of $0.1$). Predicted periods as a function of stellar parameters are obtained from analytic relations[35] for different pulsation modes (cross: second overtone; filled circle: first overtone; filled circle with a yellow dot: fundamental mode), using a colour code that depends on the surface C/O (purple: C/O $< 1$, magenta: C/O $> 1$). The labels $P_2$, $P_1$, $P_0$ are placed close to the model sequences that share the same pulsation mode (second overtone, first overtone, and fundamental, respectively).